\documentclass[useAMS,usenatbib]{mn2e}
\pdfoutput=1

\usepackage{graphicx}
\usepackage{dcolumn}
\usepackage{bm}
\usepackage{amssymb,amsmath}  
\usepackage{color}
\usepackage{hyperref}
\usepackage{lipsum}
\usepackage[usenames,dvipsnames]{xcolor}
\hypersetup{colorlinks=true,citecolor=blue,linkcolor=OliveGreen}

\newcommand{\nv}{\hat{\bf n}}

\title[The Tidal Field in the Projected Galaxy Distribution]
{Recovering the Tidal Field in the Projected Galaxy Distribution}
\author[David Alonso, Boryana Hadzhiyska, Michael A. Strauss]
{David Alonso$^1$\thanks{E-mail: david.alonso@astro.ox.ac.uk}, 
 Boryana Hadzhiyska$^2$\thanks{E-mail: boryanah@princeton.edu},
 Michael A. Strauss$^2$\thanks{E-mail: strauss@astro.princeton.edu}\\
 $^{1}$University of Oxford, Denys Wilkinson Building, Keble Road, Oxford, OX1 3RH,  UK\\
 $^{2}$Department of Astrophysical Sciences, Princeton University, Princeton, NJ 08544 USA
}

\begin{document}
  \date{\today}
  \pagerange{1--18} \pubyear{2015}
  \maketitle

\begin{abstract}
  We present a method to recover and study the projected gravitational tidal forces from a galaxy
  survey containing little or no redshift information. The method and the physical interpretation
  of the recovered tidal maps as a tracer of the cosmic web are described in detail. We first
  apply the method to a simulated galaxy survey and study the accuracy with which the cosmic web
  can be recovered in the presence of different observational effects, showing that the projected
  tidal field can be estimated with reasonable precision over large regions of the sky. We then
  apply our method to the 2MASS survey and present a publicly available full-sky map of the
  projected tidal forces in the local Universe. As an example of an application of these data we
  further study the distribution of galaxy luminosities across the different elements of the cosmic
  web, finding that, while more luminous objects are found preferentially in the most dense
  environments, there is no further segregation by tidal environment.
\end{abstract}


\section{Introduction}\label{sec:intro}
  The nature of environmental effects in structure formation is an important field of study,
  both in astrophysics and cosmology. The dependence of halo and galaxy abundances on 
  environmental density, for instance, gives rise to the bias relation linking
  the halo/galaxy distribution to the true matter density \citep{1996MNRAS.282..347M,
  1999MNRAS.308..119S} , and represents a central issue in the road to maximizing the
  amount of information that can be extracted from galaxy clustering analyses. Likewise,
  the environmental tidal forces are expected to distort the intrinsic shapes and alignments of
  galaxies, and therefore this effect must be correctly understood in order to obtain
  unbiased results from weak lensing studies \citep{2001MNRAS.320L...7C,2004PhRvD..70f3526H}.
  
  Although the study of environmental effects has traditionally been focused on the impact
  of the environmental density, in recent years many groups have studied the effect of
  other quantities, such as the morphology of the environmental density field, the local
  tidal forces or the velocity field \citep{2008MNRAS.383.1655S,2009MNRAS.396.1815F,
  2010MNRAS.409..156B,2012MNRAS.425.2049H}. Each of these quantities can be used to define a
  different classification scheme of environment types in order to describe the statistics
  of the so-called ``cosmic web'': i.e. the arangement of the matter distribution into
  interconnected structures of different dimensionality \citep{1996ApJS..103....1B}. Although
  the use of these additional observables undoubtedly furthers our understanding of
  environmental effects in structure formation, these quantities themselves may also contain
  relevant cosmological information.
  
  There has been extensive work in quantifying and understanding the properties of the cosmic
  web from N-body simulations, as well as its interplay with various intrinsic halo and galaxy
  properties \citep{2007MNRAS.375..489H,Yan2013,2014MNRAS.443.1090F,2014MNRAS.443.1274L,
  2014MNRAS.445..988N,2015MNRAS.446.1458M} and a smaller number of groups have attempted similar
  studies on galaxy survey data \citep{2010MNRAS.406.1609B,2015MNRAS.448.3665E,
  2015arXiv150906376C}. Often this is done by using the galaxy number density in redshift space as
  a proxy for the real-space matter density. However this approach entails a number of
  difficulties, such as the theoretical uncertainty in the relationship between galaxies and
  dark matter or the presence of redshift-space distortions. Furthermore, it is often difficult to
  measure accurate redshifts for a sufficiently large number of sources (e.g. very faint galaxies)
  and, although one can resort to the use of photometric redshifts, the lack of precise radial
  information precludes any attempt at an accurate reconstruction of the three-dimensional density
  field necessary to study the cosmic web. In these cases, however, there is still a significant
  amount of information encoded in the projected two-dimensional galaxy distribution, which could
  potentially be used to study the statistics of the cosmic web. In this work we present a method
  to carry out this kind of analysis, which we then implement on the Two Micron All-Sky Survey
  (2MASS hereafter \citep{2006AJ....131.1163S}), a low-redshift imaging catalog, to produce a
  full-sky map of the projected tidal forces in the local Universe.
  
  This paper is structured as follows. In section \ref{sec:theory} we present our method as a
  two-dimensional implementation of the cosmic web classification based on the tidal tensor,
  and interpret the recovered observable in terms of the projected transverse tidal forces.
  We validate this method in Section \ref{sec:sims} by implementing it on an N-body-based
  sythetic galaxy catalog, and devise a technique to deal with an incomplete sky coverage.
  The implementation of the method on the 2MASS survey is presented in Section \ref{sec:2mass}.
  As a proof of concept we also use the produced maps of the projected tidal field to study
  the dependence of the luminosity function on the tidal classification of the environment.
  Our conclusions are presented in Section \ref{sec:discussion}.

\section{The 2D tidal tensor and the projected cosmic web}\label{sec:theory}
  \subsection{The three-dimensional tidal tensor}\label{ssec:th_3d}
    One of the most popular methods used in the literature to study the properties
    of the cosmic web is through the structure of the gravitational tidal forces
    \citep{1970Ap......6..320D,2007MNRAS.375..489H,2009MNRAS.396.1815F}.  
    The action of these forces on an extended body of size ${\bf l}$ stretches or
    contracts it along different directions based on the structure of the Hessian
    of the gravitational potential $\bar{\Phi}$, also called the tidal tensor field:
    \begin{equation}
      \ddot{l}_i=-l_j\,\partial_j\partial_i\bar{\Phi}\,.
    \end{equation}
    The tidal tensor is symmetric, and therefore can always be diagonalised at
    any point in space by performing a three-dimensional rotation. The eigenvalues
    of the tidal tensor therefore inform us about the strength of the tidal forces
    in three independent orthogonal directions, and their sign can be used to classify
    four different types of environments. In the standard cosmic web classification,
    at a point in space in which all the eigenvalues are positive, extended objects will
    be compressed in all directions, and such a point is classified as a \emph{knot}.
    On the other end, objects in a region where the tidal field has all-negative eigenvalues
    will be stretched in all directions, and the region is classified as a \emph{void}.
    The intermediate cases correspond to filaments (two positive and one negative
    eigenvalues) and sheets (one positive and two negative eigenvalues).
    
    Note that, even though this nomenclature alludes to the geometrical or morphological
    properties of these structures, the method is entirely based on the properties of
    the tidal field, and thus is dynamical in nature. This is different from the
    alternative approach of separating distinct\ elements of the cosmic web in terms
    of the morphology of the density field (e.g. \citet{2010MNRAS.406.1609B}). An added
    value of dynamical prescriptions is the direct physical interpretation of the
    resulting structures in terms of contracting and expanding directions, which can
    have a direct impact on the physics of galaxy formation. Other similar
    methods based on the tidal tensor or the velocity shear tensor 
    have been proposed in the literature \cite{2008MNRAS.383.1655S,2010MNRAS.406.1609B,
    2012MNRAS.425.2049H,2013MNRAS.428.2489L} following a similar rationale.

    Here we will adhere to the formalism used in \cite{2009MNRAS.396.1815F,2015MNRAS.447.2683A}.
    For simplicity we will work with a rescaled version of the Newtonian potential:
    $\Phi\equiv\bar{\Phi}/(4\pi G\bar{\rho})$, for which the Poisson equation is simply
    $\nabla^2\Phi=\Delta$, where $\Delta$ is the matter overdensity field. We then define
    the tidal tensor field as $T_{ij}=\partial_i\partial_j\Phi$, so that
    $\Delta={\rm Tr}(\hat{T})$, and we will classify the environment according to the
    number of eigenvalues $\alpha$ above a given threshold $\Lambda_{\rm th}$ (not
    necessarily $\Lambda_{\rm th}=0$). The standard approach to compute the tidal tensor in
    three-dimensional datasets is to first estimate the gravitational potential $\Phi$ by solving
    Poisson's equation in Fourier space, and then differentiate it (also in Fourier space) to
    compute its Hessian. Thus, the Fourier transforms of the tidal tensor and the
    three-dimensional density field are related through
    \begin{equation}
      T_{ij}({\bf k})=\frac{k_ik_j}{k^2}\Delta({\bf k}).
    \end{equation}
    The density field used for these analyses is usually smoothed down to a given scale $R_s$,
    either to mitigate shot-noise, filter out non-linear effects or in order to study the
    scale-dependence of the resulting tidal field. We will do the same in the 2-dimensional
    case.

  \subsection{The 2D tidal tensor}\label{ssec:th_2d}
    \subsubsection{Definition}\label{sssec:th_2d_def}
      The formalism introduced above is straightforward to implement in
      an N-body simulation, and methods have been devised to use it also in spectroscopic
      galaxy catalogues (e.g. \cite{2010MNRAS.409..156B,2010MNRAS.406..320C,
      2015MNRAS.448.3665E}), where three-dimensional positions can be accurately determined
      for all galaxies (at least up to the effect of peculiar velocities). However, determining
      accurate spectroscopic redshifts for individual galaxies is a very time-consuming
      operation, and often in astronomy we are forced to make do with datasets for which
      radial positions are very poorly measured, as is the case for photometric redshift
      surveys, or even completely unknown. While a large amount of information is lost in the
      absence of accurate radial positions, a sizeable portion of it still remains encoded in
      the projected angular distribution of galaxies. The method presented here is intended to
      enable the study of environmental tidal forces in the matter distribution in these cases.
      
      The idea behind this method is to use a straightforward dictionary between
      three-dimensional and two-dimensional projected quantities. A proxy of the transverse
      components (i.e. perpendicular to the line of sight) of the projected
      tidal tensor is computed from the projected density field using the following
      prescription:
      \begin{itemize}
        \item The direct observable in a projected dataset is the projected overdensity
              $\delta(\nv)$: the fluctuations in the angular number density of galaxies
              with respect to the mean. This is related to the three-dimensional overdensity
              field $\Delta({\bf x})$ through a line-of-sight projection:
              \begin{equation}
                \delta(\nv)=\int_0^\infty d\chi\,w(\chi)\,\Delta_s(\chi\nv),
              \end{equation}
              where $\chi$ is the comoving radial distance, $w(\chi)$ is the survey selection
              function and $\Delta_s({\bf x})$ is the redshift-space three-dimensional
              overdensity field. Here and in what follows we will denote all projected
              quantities using the lower-case version of the symbols used for the analogous
              three-dimensional objects.
              
              As mentioned in the previous section, we filter out the smallest scales of the
              density field to mitigate shot-noise effects and non-linearities. For this we
              will use a Gaussian smoothing kernel defined by its standard deviation $\theta_s$.
        \item We define the \emph{2D potential} $\phi$ as the solution to Poisson's equation
              on the sphere with $\delta$ as a source:
              \begin{equation}\label{eq:phi}
                \nabla^2_{\nv} \phi\equiv\delta,
              \end{equation}
              where $\nabla^2_{\nv}\equiv \partial_\theta^2+\partial_\varphi^2/\sin^2\theta+
              \cot\theta\partial_\theta$ is the covariant Laplacian on the sphere, and $\theta$
              and $\varphi$ are the elevation and azimuth spherical coordinates respectively.
              Note that $\phi$ thus defined is not the same as the projected potential
              $\tilde{\phi}$, given by
              \begin{equation}
                \tilde{\phi}(\nv)\equiv\int_0^\infty d\chi\,w(\chi)\,\Phi(\chi\nv).
              \end{equation}
              We will discuss these differences in more detail in Sections
              \ref{sssec:th_2d_phys}, \ref{ssec:2d23dsim} and Appendix \ref{app:th_fullsky}.
        \item The \emph{2D tidal tensor} $t_{ab}$ is then defined as the covariant Hessian of
              the 2D potential, $t_{ab}\equiv H_{ab}\,\phi$, where the covariant Hessian
              operator is given by
              \begin{equation}\label{eq:cov_hess}
                \hat{H}\equiv\left(
                \begin{array}{ccc}
                 \partial_\theta^2 & \partial_\theta(\partial_\varphi/\sin\theta) \\
                 \partial_\theta(\partial_\varphi/\sin\theta) &
                 \partial_\varphi^2/\sin^2\theta+\cot\theta\partial_\theta
                \end{array}
                \right).
              \end{equation}
      \end{itemize}
      
      The procedure outlined above is nothing but a direct analogy with what is done to obtain the
      three-dimensional tidal tensor: find the potential by solving Poisson's equation using
      Fourier methods and then compute the second derivatives of that potential. Although this
      is a simple way to define the 2D tidal tensor, we must first understand the physical
      interpretation of the object thus computed. We do so in the next section.
    
    \subsubsection{Physical interpretation of the 2D tidal tensor}\label{sssec:th_2d_phys}
      The physical interpretation of the 2D tidal tensor introduced in the previous section is
      most easily understood in the flat-sky approximation. In this case, the projected
      overdensity is related to the three-dimensional one through:
      \begin{equation}
        \delta({\bf x})\equiv\int dz\,w(z)\Delta_s({\bf x},z).
      \end{equation}
      Here ${\bf x}=(x,y)$ are the coordinates perpendicular to the line of sight, and we have
      chosen $z$ to be the radial coordinate. $\Delta_s({\bf x},z)$ is the three-dimensional
      overdensity field in redshift-space, and $w(z)$ is the radial selection function.

      It is easy to relate $\delta({\bf x})$ to the Fourier transform of the 3D matter
      overdensity field:
      \begin{equation}\label{eq:3dto2d}
         \delta({\bf x})=\int\frac{dk^2}{2\pi}e^{i{\bf k}{\bf x}}
                         \int dq\,w(q)\,b\left[1+\beta\frac{q^2}{k^2+q^2}\right]
                         \Delta({\bf k},q),
      \end{equation}
      where ${\bf k}$ and $q$ are the components of the wave vector perpendicular and
      parallel to the line of sight respectively, and $w(q)$ is the Fourier
      transform of the selection function
      \begin{equation}
        w(q)\equiv\int \frac{dz}{\sqrt{2\pi}}\,w(z)e^{i\,qz}.
      \end{equation}
      The factors $b$ and $\beta$ in Eq. \ref{eq:3dto2d} account for the galaxy bias
      and linear redshift-space distortions (i.e. $\beta\equiv f/b$, where $f\equiv
      d\log D/d\log a$ is the linear growth rate).

      According to the definition used in the previous section, the 2D tidal tensor
      $\hat{t}$ and its three-dimensional version $\hat{T}$ along the two transverse
      directions are related to the projected and three-dimensional density fields in
      Fourier space respectively through
      \begin{equation}
        t_{ab}\equiv\frac{k_ak_b}{k^2}\delta({\bf k}),\hspace{12pt}
        T_{ab}\equiv\frac{k_ak_b}{k^2+q^2}\Delta({\bf k},q),
      \end{equation}
      and thus, they are related to each other through
      \begin{equation}\label{eq:tab_flat}
        t_{ab}({\bf x})=b\,\int\frac{dk^2}{2\pi}e^{i{\bf k}{\bf x}}
        \int dq\,\omega(q,k)\,T_{ab}({\bf k},q),
      \end{equation}
      where we have defined the modified selection function:
      \begin{equation}
        \omega(q,k)\equiv w(q)\,\left[1+(1+\beta)\frac{q^2}{k^2}\right],
      \end{equation}

      On the other hand, the transverse components of the three-dimensional
      tidal tensor projected along the line of sight are given by
      \begin{equation}\label{eq:ttab_flat}
        \tilde{t}_{ab}({\bf x})\equiv\int\frac{dk^2}{2\pi}e^{i{\bf k}{\bf x}}
        \int dq\,w(q)\,T_{ab}({\bf k},q).
      \end{equation}
      Comparing Equations \ref{eq:tab_flat} and \ref{eq:ttab_flat} we can see that the
      differences between both quantities are fully encapsulated in the different
      selection functions $w$ and $\omega$\footnote{as well as the multiplicative
      galaxy bias factor $b$, due to the fact that the true tidal field is caused
      by the total matter density.}.
      
      Typically, the selection function $w$ of any survey will have a characteristic
      radial width $l_z$, and therefore its Fourier transform will have support over
      a range of scales $q\lesssim 1/l_z$ (e.g., the Fourier transform of a Gaussian
      selection function with variance $l_z$ is a Gaussian with
      variance $1/l_z$). Since $\omega$ and $w$ differ significantly only for
      values of $q\gtrsim k$ (we assume that the RSD parameter $\beta$ is $O(1)$),
      this implies that, as long as we focus only on angular scales $k\gtrsim1/l_z$,
      the 2D tidal tensor $t_{ab}$ and the projected tidal tensor $\tilde{t}_{ab}$
      will be proportional to each other to a very good approximation.
      
      Hence, for sufficiently wide window functions, $t_{ab}$ can be safely
      interpreted on all scales of interest as the magnitude of the tidal forces in
      the transverse directions averaged along the line of sight over the survey
      selection function. A more rigorous and quantitative proof of this result using
      a full-sky formalism is presented in Appendix \ref{app:th_fullsky}, and we demonstrate
      it in practice in Section \ref{ssec:2d23dsim}.

    \subsubsection{Classification of the projected cosmic web}\label{sssec:th_classify}
      Bearing in mind the physical interpretation of the 2D tidal tensor, we can now
      justify an environmental classification based on it. We thus define three
      different types of environments in terms of the number of eigenvalues of the 2D tidal
      tensor larger than a given eigenvalue threshold $\lambda_{\rm th}$. We chose to
      retain the names \emph{knots} and \emph{voids} to denote regions in which both or
      none of the eigenvalues exceed the threshold respectively, and we label any
      region in which only one of the eigenvalues is found above the threshold as a
      \emph{nexus}. In what follows we will order the two eigenvalues of the 2D
      tidal tensor have so that $\lambda_1\geq\lambda_2$, and therefore our
      prescription for the tidal classification reads:
      \begin{enumerate}
        \item {\bf Void}: all eigenvalues below the threshold
                           ($\lambda_1\leq\lambda_{\rm th}$).
        \item {\bf Nexus}: only 1 eigenvalue above the threshold
                           ($\lambda_2\leq\lambda_{\rm th}<\lambda_1$).
        \item {\bf Knot}: all eigenvalues above the threshold
                           ($\lambda_{\rm th}\leq\lambda_2$).
      \end{enumerate}
      
      This formalism has one free parameter: the eigenvalue threshold $\lambda_{\rm th}$.
      In the three-dimensional case, several prescriptions have been proposed in the
      literature to choose a value for the analogous parameter $\Lambda_{\rm th}$.
      A choice of $\Lambda_{\rm th}=0$ would separate different regions based purely on
      the direction of the tidal forces. This prescription would assume that gravitational
      collapse is underway along a given direction even if the eigenvalue is only
      infinitesimally positive, although in this case collapse would only occur after a very
      long time. This prescription thus produces a tidal classification in which voids
      occupy only about $\sim20\%$ of the volume, in striking contrast with the visual
      impression from redshift surveys and N-body simulations that most of the volume is
      actually empty. A choice of $\Lambda_{\rm th}>0$ would therefore only regard a given
      direction as ``collapsing'' if the tidal forces are sufficiently strong, and would
      give rise to a tidal classification in which the abundance of voids better matches
      our intuitive expectations. The spherical collapse model would suggest an appropriate
      value for $\Lambda_{\rm th}\sim O(1)$ \cite{2009MNRAS.396.1815F}, which would actually
      classify a large fraction of regions with overdensities $\delta\gtrsim1$ as voids.
      This is partly due to the failure of the spherical collapse model to describe the
      physics of anisotropic collapse, and therefore an intermediate value of
      $\Lambda_{\rm th}$ is sometimes chosen in order to produce the correct visual
      impression of the cosmic web classification. 

      In order to avoid this arbitrariness we have followed a prescription similar to
      the one proposed by \cite{2015MNRAS.448.3665E}: for different values of the
      eigenvalue threshold we calculate the number of galaxies in the survey located
      in the three different environmental types, and we choose the value of
      $\lambda_{\rm th}$ that most equally divides the galaxy population among the
      different types, thus minimizing the statistical uncertainty when studying the
      statistics of the galaxy population in all the environments simultaneously.
      The exact procedure we used to select a value for $\lambda_{\rm th}$ is the
      following: for the three environment types: ($\alpha=(0,1,2)$) we compute
      the fraction of galaxies in it: $F_\alpha=N_\alpha/N_{\rm total}$. We then
      compute the root-mean-square deviation in these fractions as:
      \begin{equation}
        \Sigma_F=\frac{1}{3}\sqrt{\sum_{\alpha=0}^{2}
        \left(F_\alpha-\frac{1}{3}\right)^2},
      \end{equation}
      and select the value of $\lambda_{\rm th}$ that minimizes $\Sigma_F$. The actual
      value of $\lambda_{\rm th}$ depends on the smoothing scale used as well as on the
      galaxy population under study. As shown in Section \ref{ssec:2m_results}, for
      our fiducial smoothing scale of $1^\circ$ and the 2MASS sample, the optimal value
      is $\lambda_{\rm th}=0.05$. Note that, although this criterion maximizes the
      statistics in the three environments simultaneously, it is not necessarily the
      optimal choice in order to enhance possible tidal effects in the galaxy
      distribution.
      
\section{Testing the method on simulated datasets}\label{sec:sims}
  \subsection{Simulations}\label{ssec:sim_description}
    We have first tested the method to estimate the 2D tidal tensor on a simulated galaxy
    catalog in order to rigorously verify the different systematic effects that could
    contaminate the measurements in the real data. The simulated data will
    also allow us to evaluate the agreement of these measurements with our theoretical
    expectations.

    Our target galaxy catalog is the 2MASS survey, described in Section \ref{ssec:2m_describe},
    and therefore we generated the simulated catalog to match 2MASS in terms of number density,
    clustering amplitude and redshift and magnitude distributions. The base of our simulated
    survey is a dark-matter-only N-body
    simulation, produced by the hybrid TreePM code {\tt Gadget-2} \cite{2005MNRAS.364.1105S}.
    It was run on a cubic box of size $L_{\rm box}=700\,{\rm Mpc}/h$ containing $1024^3$
    particles with a particle mass $m_p=2.7\times10^{10}\,M_\odot/h$. This mass resolution was
    necessary to populate the dark-matter haloes with galaxies matching the redshift and
    magnitude distributions of 2MASS. The simulation was run assuming a flat $\Lambda$CDM
    cosmology with cosmological parameters $(\Omega_M,\Omega_b,h,\sigma_8,n_s)=
    (0.3,0.05,0.7,0.8,0.96)$, in approximate agreement with \cite{2014A&A...571A..16P}. The
    initial conditions for the simulation were generated at redshift $z_{\rm ini}=49$ using
    second-order Lagrangian perturbation theory with a matter transfer function computed by
    the Boltzmann code {\tt CAMB} \cite{2000ApJ...538..473L} for the cosmological parameters
    above.

    A comoving snapshot of the simulation at redshift $z=0.1$ (the median redshift of
    2MASS) was used to generate the mock galaxy catalog. First, a halo catalog was generated
    using a Friends-of-Friends (FoF) code\footnote{The halo finder is publicly available
    and can be found at \url{https://github.com/damonge/MatchMaker}.} with a linking length
    of $b=0.2$ in units of the mean interparticle distance of the simulation. All haloes
    containing 5 or more particles were included in the catalog. Although it is not clear
    whether FoF groups with such a small number of particles can describe virialized structures
    accurately, the main aim of this catalog is not to study the galaxy-halo relationship
    accurately, but rather to produce a galaxy sample that matches the statistical properties
    of 2MASS, which we have managed to accomplish. Furthermore, haloes with masses below 5
    particles are only needed to reach survey completeness at the smallest redshifts
    ($z\lesssim0.03$), and therefore the impact of the galaxies populating those haloes on the
    statistics of the overall sample is almost negligible.
    
    Haloes were populated with galaxies following a simplified version of the hybrid method
    proposed in \cite{2015MNRAS.447..646C}, combining Halo Abundance Matching (HAM)
    \cite{2004MNRAS.353..189V} and Halo Occupation Distribution (HOD)
    \cite{2000MNRAS.318.1144P,2002ApJ...575..587B} techniques. All haloes with masses larger
    than a threshold $M_{\rm min}$ were assigned a single central galaxy, and a
    mass-dependent average number of satellite galaxies given by
    \begin{equation}
      \langle N_{\rm sat}(M)\rangle=\left(\frac{M}{M_1}\right)^\beta,
    \end{equation}
    and all haloes below $M_{\rm min}$ were left empty.
    The actual number of satellite galaxies assigned to each halo was drawn from a
    Poisson distribution with this mean. In order to assign luminosities to each
    galaxy following the 2MASS luminosity function we first related halo masses to
    luminosities by matching the cumulative luminosity function and the cumulative
    galaxy number density as a function of halo mass. Specifically, halo masses ($M_h$)
    were related to $K_s$-band luminosities ($L_K$) by solving the equation:
    \begin{equation}
      \int_{L_K}^\infty dL\,\frac{dn_g}{dL}=\int_{M_h}^\infty dM\,n(M)
      \left[1+\langle N_{\rm sat}(M)\rangle\right]
    \end{equation}
    where $dn_g/dL$ is the $K_s$-band luminosity function and $n(M)$ is the halo
    mass function. For the mass function we chose the parametrization by
    \cite{2001MNRAS.321..372J}, which we found matches the halo mass distribution
    in our simulation to good accuracy. For the luminosity function we used the
    measurements of \cite{2014JCAP...10..070A} at $K_s<13.5$, in which $dn_g/dL$ is
    modelled as a Schechter function with parameters
    $\phi^*=1.10\times10^{-2}\,({\rm Mpc}/h)^{-3}$, $M_K^*=-23.52+5\log(h)$ and
    $\alpha=-1.00$. 
    
    Once the $M_h-L_K$ relation is found, central galaxies are assigned the luminosity
    corresponding to the host halo mass, and satellites are given a luminosity
    drawn from the satellite luminosity function, given in its cumulative form by
    \begin{equation}
      n_{\rm sat}(>L_K)\equiv\int_{L_K}^\infty dL\,\frac{dn_g}{dL}-
      \int_{M_h(L_K)}^\infty dM\,n(M).
    \end{equation}
    This algorithm guarantees that the resulting galaxy sample will follow the
    input luminosity function. Once intrinsic luminosities were assigned, the apparent
    $K_s$ magnitude was computed for each galaxy using Eq. \ref{eq:rel2abs}, and a cut
    $K_s\geq13.9$ (corresponding to the 2MASS completeness limit) was imposed.
    
    The three free parameters of this method, $M_{\rm min}$, $M_1$ and $\beta$
    were fixed by matching the amplitude of the angular power spectrum in the
    simulation to that of the 2MASS data for three galaxy samples with different limit magnitudes:
    $K_s<13.5$, $K_s<13.8$ and $K_s<13.9$. We found the combination
    $\log_{10}[M_{\rm min}/(M_\odot/h)]=10.7$, $\log_{10}[M_1/(M_\odot/h)]=13.5$ and $\beta=1.4$
    to yield a good agreement with the data in terms of clustering amplitude, number
    density and redshift distribution.
    \begin{figure}
      \centering
      \includegraphics[width=0.49\textwidth]{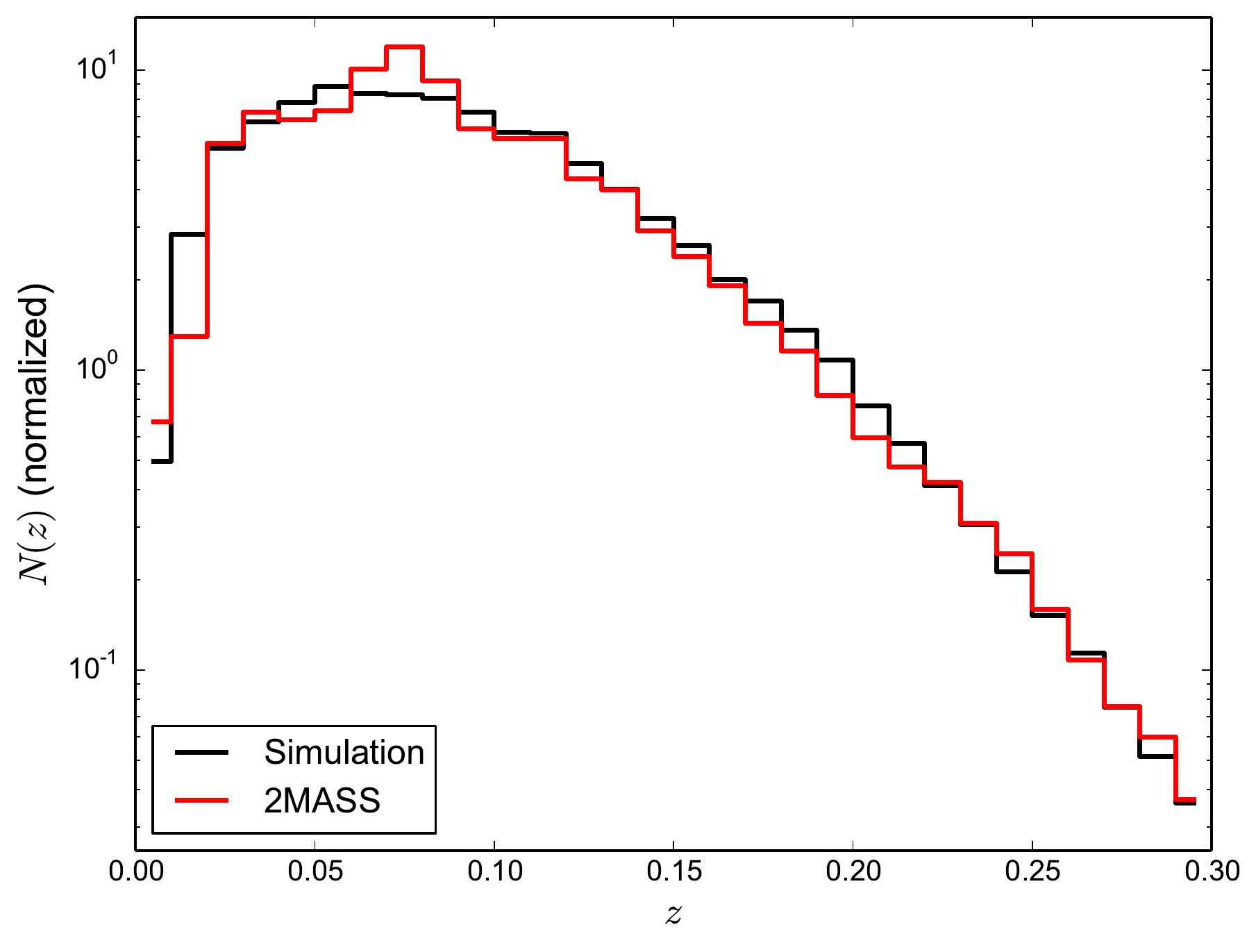}
      \caption{Redshift distribution of the simulated galaxy catalog (black) and the sample of
        2MASS galaxies with $K_s<13.9$ used in this analysis (red). The latter was estimated from
        the redshifts of the complete spectroscopic sample in the northern galactic hemisphere,
        comprising $\sim113000$ objects (see Section \ref{sssec:2m_lfun}). Both histograms have
        been normalized to unit area.}
      \label{fig:sim_nz}
    \end{figure}
    \begin{figure*}
      \centering
      \includegraphics[width=0.49\textwidth]{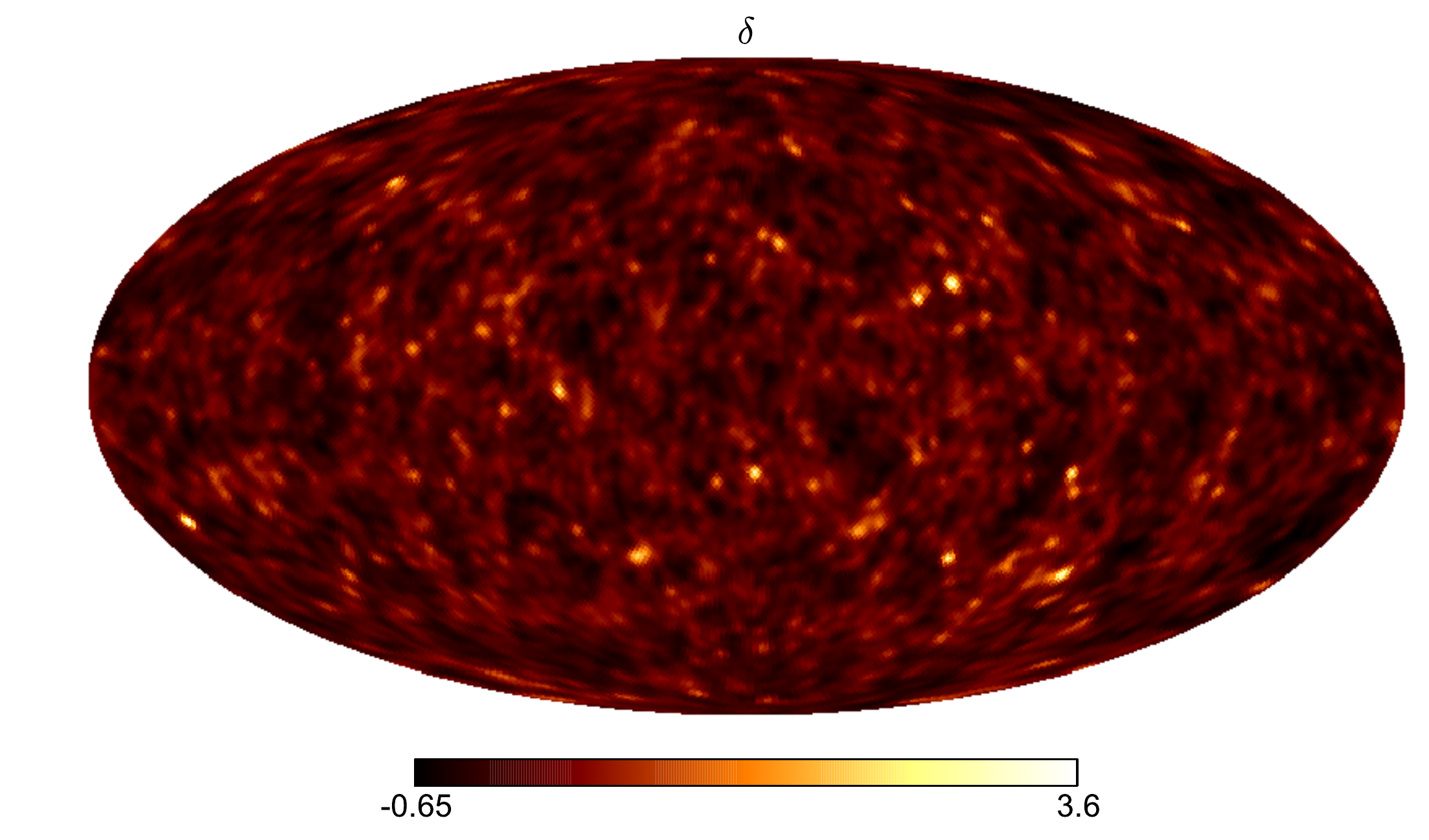}
      \includegraphics[width=0.49\textwidth]{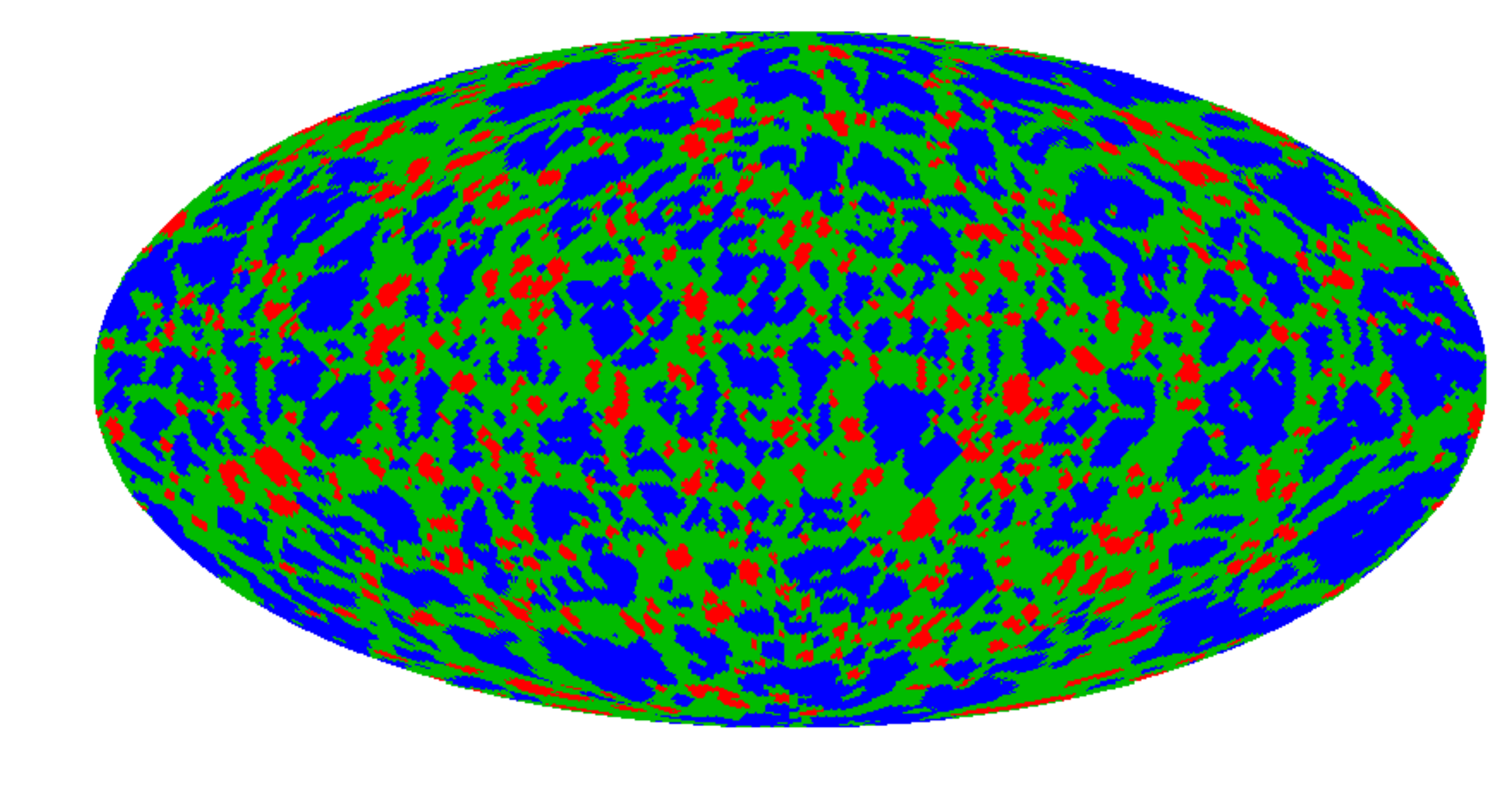}
      \includegraphics[width=0.49\textwidth]{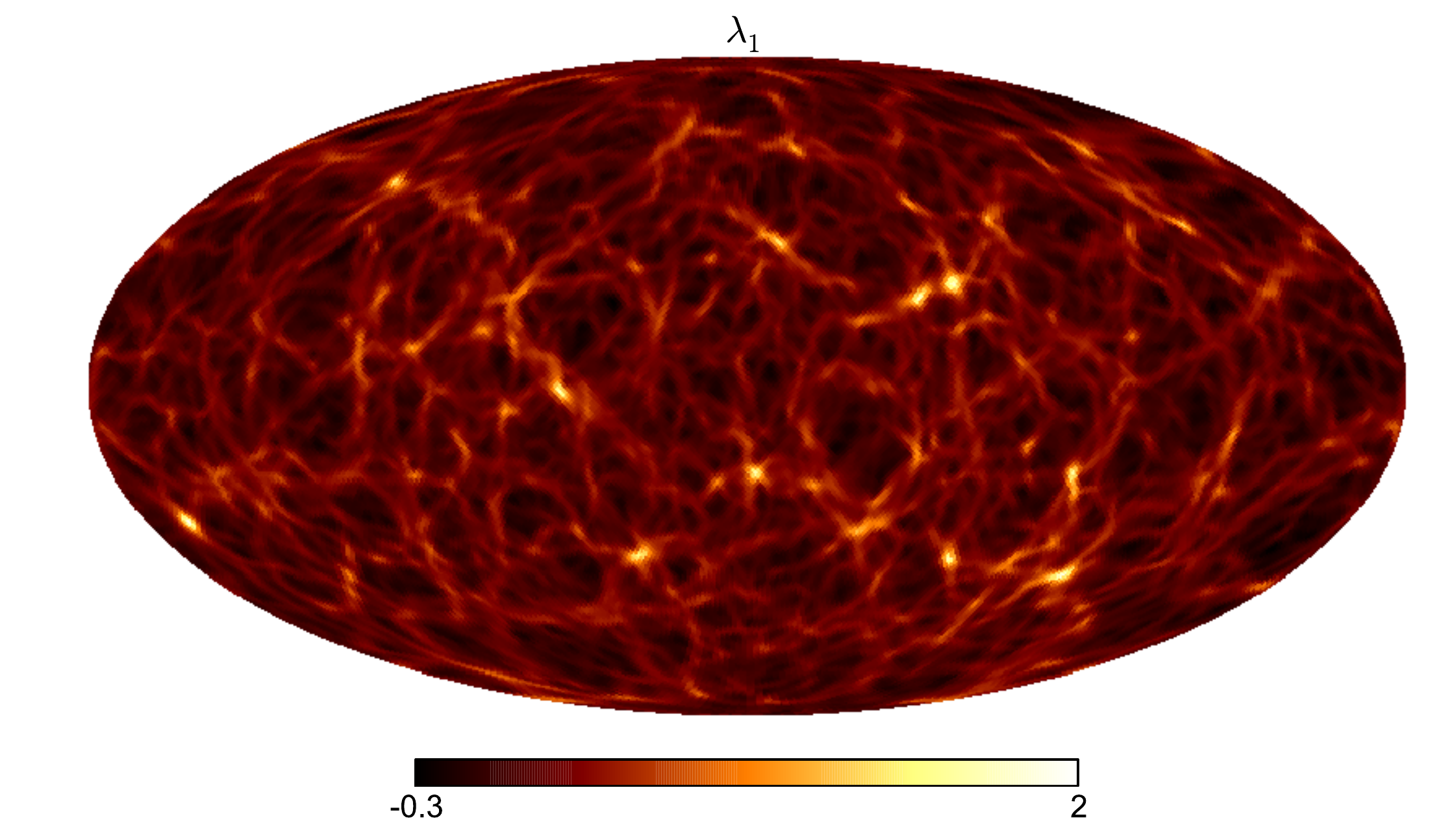}
      \includegraphics[width=0.49\textwidth]{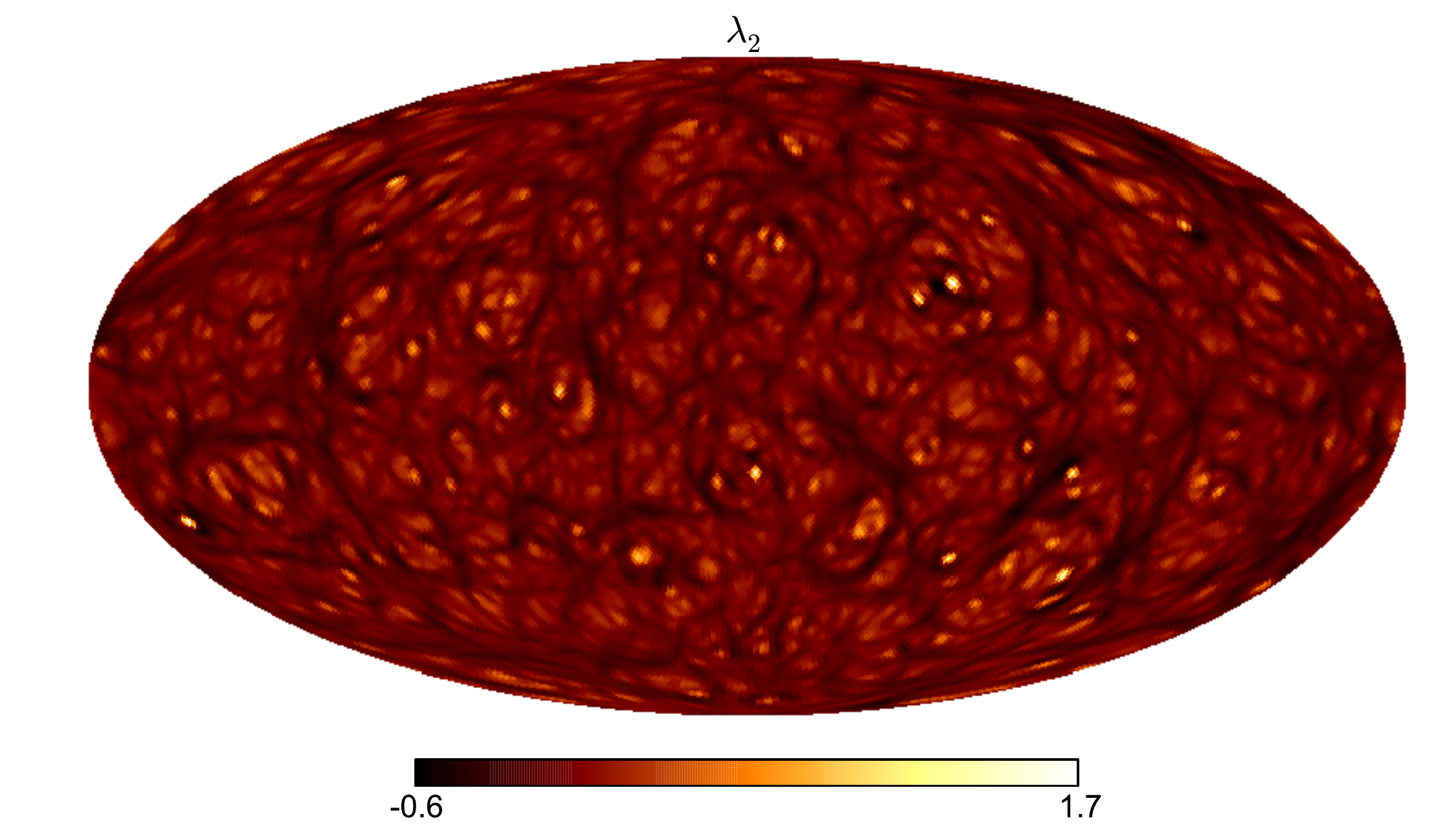}
      \caption{{\sl Top left panel:} density field of the simulated galaxy catalog smoothed with a
         $1^{\circ}$ Gaussian kernel. {\sl Top right panel:} environment classification
         for the fiducial threshold $\lambda_{\rm th}=0.05$, with knots, nexuses and voids shown in
         red, green and blue respectively. {\sl Bottom panels:} full-sky maps of the two eigenvalues
         of the 2D tidal tensor.}
      \label{fig:sim_dens}
    \end{figure*}
    
    Finally, we must note that the size of the simulation box is not quite large enough to
    encompass the whole volume covered by 2MASS. In order to achieve the required volume we
    replicated the box once in each of the three dimensions, making use of the periodic boundary
    conditions of the simulation. This implies that our simulated catalog lacks all clustering
    modes larger or similar to the size of the simulation box ($700\,{\rm Mpc}/h$), which is
    irrelevant for the scales used in the comparison of the mock catalog with our 2MASS sample.
    The redshift distribution of our simulated catalog is shown in Fig. \ref{fig:sim_nz} together
    with the corresponding one for 2MASS, extracted from the spectroscopic sample described in
    Section \ref{sssec:2m_lfun}.

  \subsection{Statistics of the projected cosmic web}\label{ssec:sim_ideal}
    We first study the 2D tidal tensor and the projected cosmic web classification 
    for the sample of galaxies in our simulated catalog matching the fiducial sample
    used in the analysis of the 2MASS data, comprised of all galaxies with
    apparent magnitudes $K_s<13.9$. For this sample we carry out the steps
    outlined in Section \ref{sssec:th_2d_def}.
    \begin{figure*}
      \centering
      \includegraphics[width=0.49\textwidth]{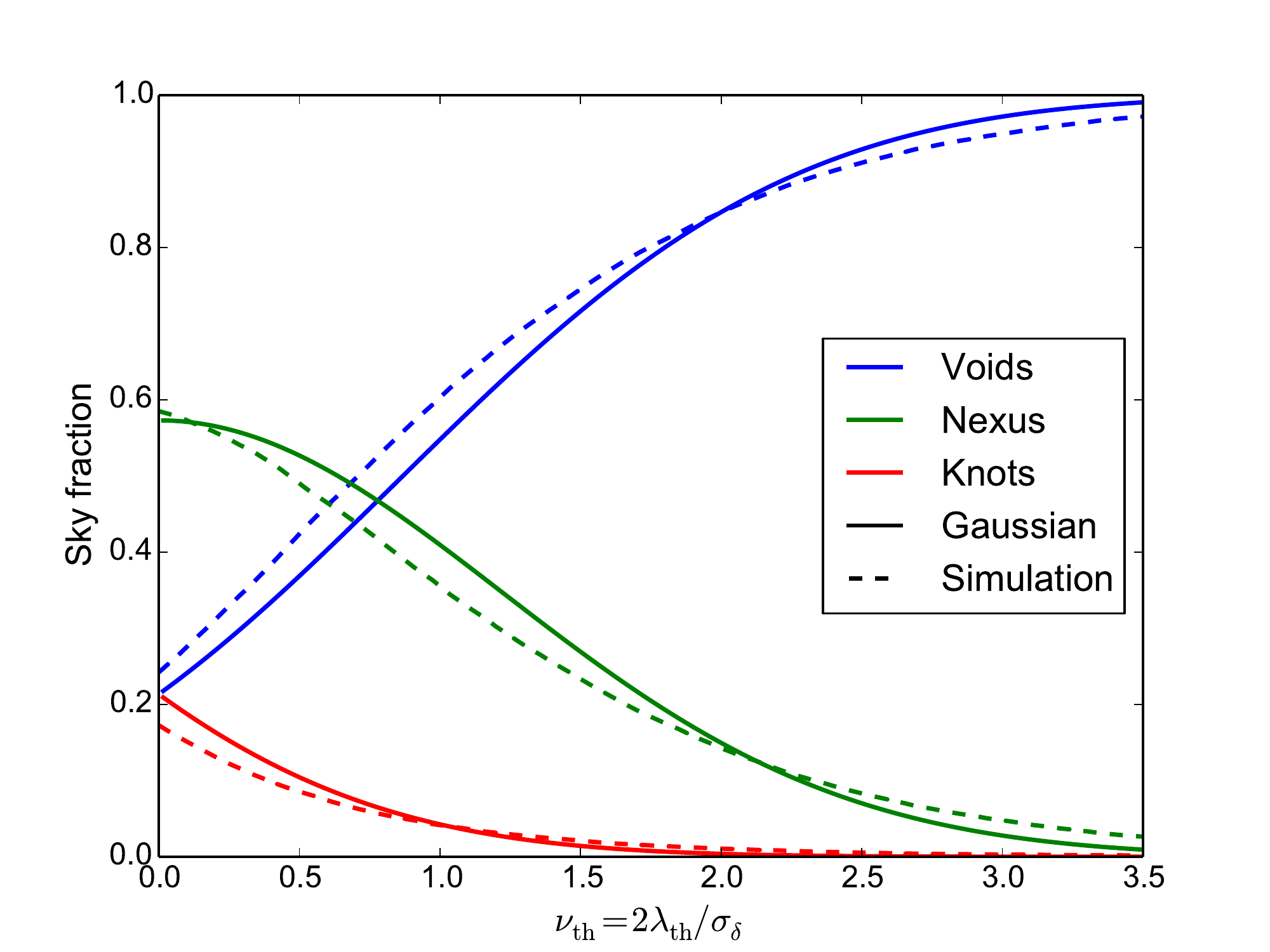}
      \includegraphics[width=0.49\textwidth]{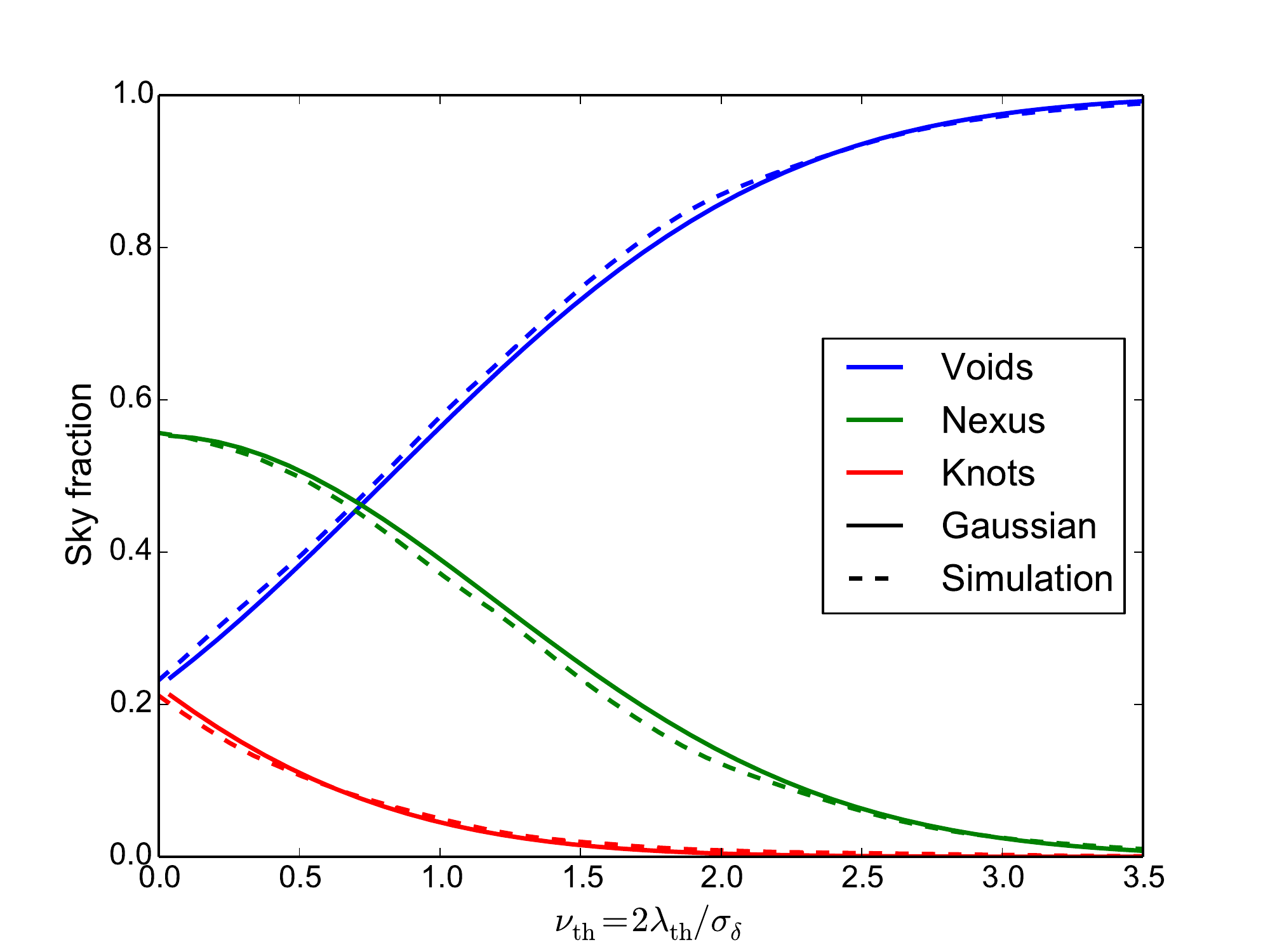}
      \caption{Gaussian prediction for the sky fraction each environment occupies
        as a function of the ratio of the eigenvalue threshold and the standard
        deviation of the projected overdensity field (solid), compared with the sky
        fractions in the simulated data (dashed). The smoothing
        angle is $\theta_{\rm sm}=1^{\circ}$ for the left panel and $\theta_{\rm sm}
        =5^{\circ}$ for the right one. The effects of non-linearities are significantly
        reduced for the larger smoothing scale, and the measured sky fractions
        agree better with the Gaussian prediction.}
      \label{fig:sim_gauss}
    \end{figure*}

    \begin{figure*}
      \centering
      \includegraphics[width=0.49\textwidth]{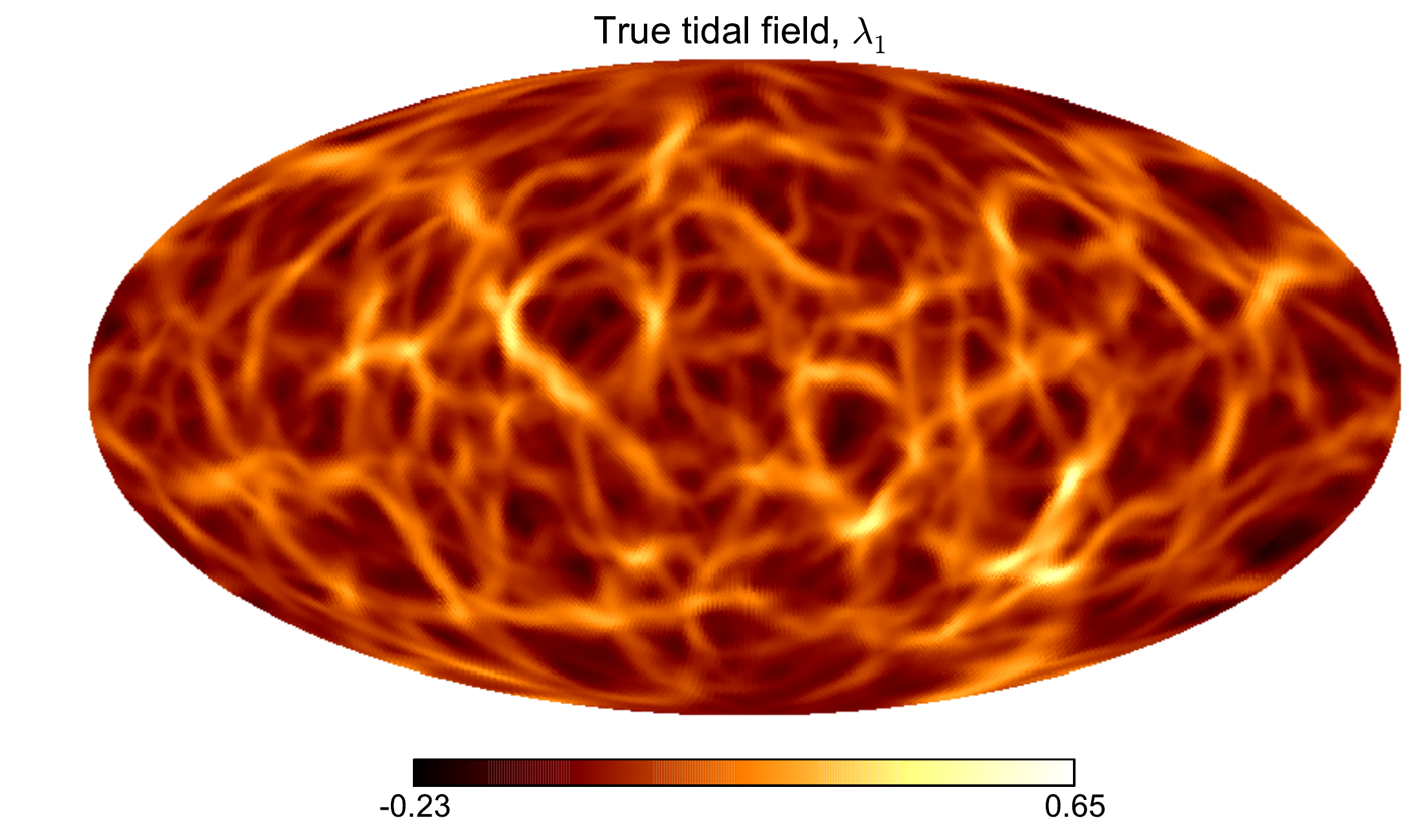}
      \includegraphics[width=0.49\textwidth]{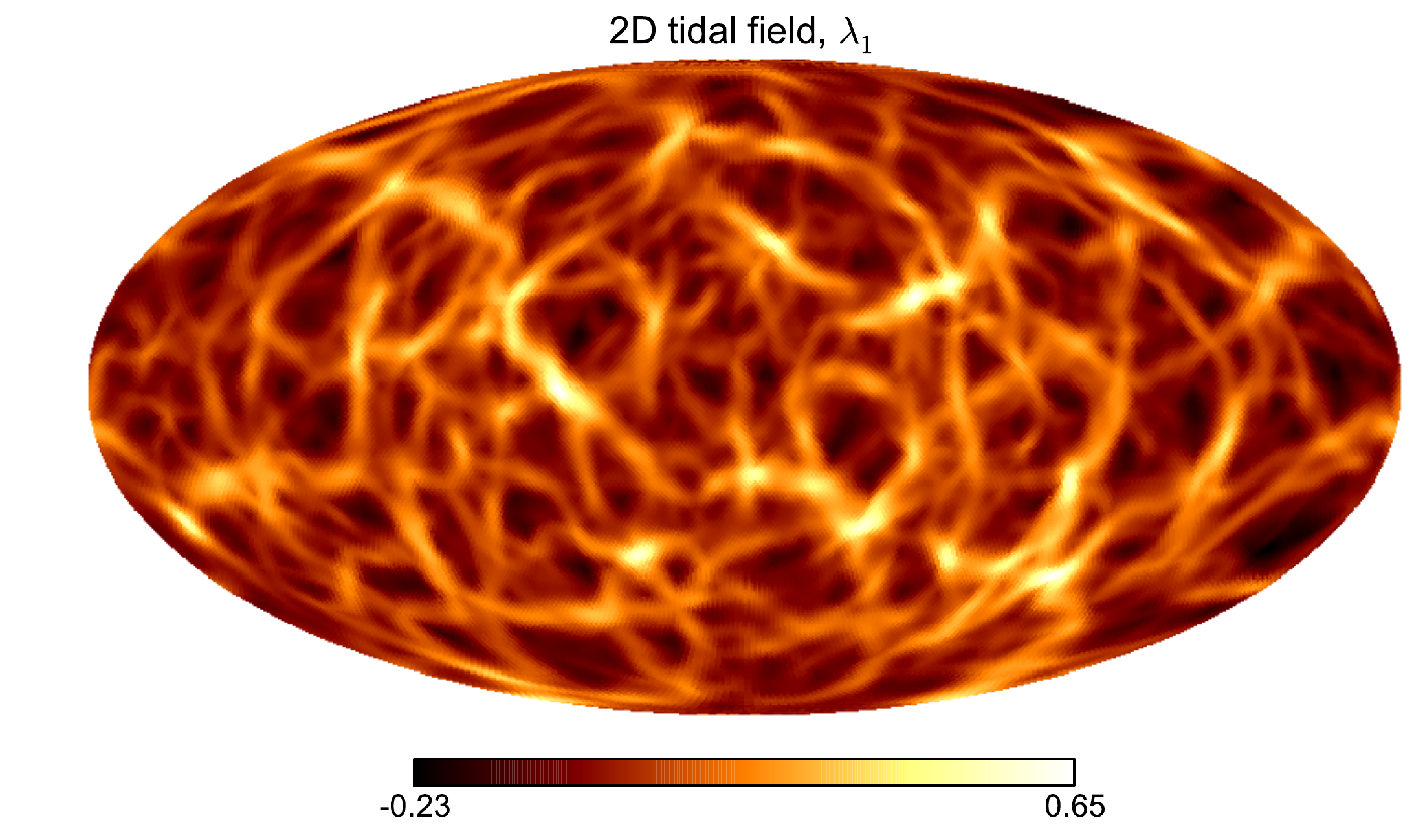}
      \caption{{\sl Left panel:} principal eigenvalue of the true projected tidal field
      in our N-body simulation. {\sl Right panel:} principal eigenvalue of the 2D tidal
      field measured from the corresponding mock galaxy catalog. A Gaussian smoothing
      kernel with $\theta_s=2^\circ$ was used in both cases. As discussed in Section
      \ref{sssec:th_2d_phys}, the 2D tidal field can be interpreted as a biased version
      of the true tidal field for most scales of interest.}
      \label{fig:eig_2dvs3d}
    \end{figure*}
    \begin{enumerate}
      \item Throughout the analysis we use the HEALPix pixelization scheme
        \cite{2005ApJ...622..759G} with a resolution parameter $N_{\rm side}=64$,
        corresponding to pixels with an area of $\sim0.84\,{\rm deg}^2$. Given
        the number density of sources in 2MASS, higher resolution parameters would
        yield an estimate of the density field overly dominated by shot noise.
      \item We compute the overdensity field in the full sky by counting the
        number of galaxies in each pixel $N_p$ and dividing by the average
        number of galaxies per pixel $\bar{N}$. The field in pixel $p$ is then given
        by:
        \begin{equation}
          \delta_p = \frac{N_p}{\bar{N}} - 1.
        \end{equation}
      \item Since our method to compute the 2D tidal tensor involves the numerical
        differentiation of the 2D potential $\phi$, in order to suppress the numerical
        noise in the computation of those derivatives we first smooth the overdensity
        field using a Gaussian smoothing kernel, with standard deviation
        $\theta_{\rm sm}=1^{\circ}$ and $5^{\circ}$. At the median redshift of our
        simulation ($\bar{z}\sim0.08$) these angles correspond to physical scales of
        4.4 and 31.2 ${\rm Mpc}/h$ respectively. The use of different smoothing scales
        also allows us to study the properties of the cosmic web as a function of scale,
        which will be useful in order to compare our results with the linear theory
        outlined in Appendix \ref{app:th_gauss}.
      \item The 2D potential $\phi$, as described by Eq. \ref{eq:phi}, is computed
        from the smoothed density field $\delta$ by solving Poisson's equation on
        the sphere. This is trivially done in harmonic space, since the harmonic
        coefficients of the two quantities are proportional to each other:
        \begin{equation}
          \phi_{\ell m} = -\frac{\delta_{\ell m}}{\ell(\ell+1)}.
        \end{equation}
      \item The 2D tidal tensor is then computed by differentiating the 2D potential
        as in Eq. \ref{eq:cov_hess}. The covariant Hessian was computed using the
        routines provided by the HEALPix {\tt python} package {\tt healpy}
        \footnote{\url{https://healpy.readthedocs.org/en/latest/}}, which
        perform the derivatives in harmonic space. The estimated tidal tensor in
        each pixel is then diagonalized, and the values of the two eigenvalues are
        used to classify each pixel as belonging to one of the three environments
        defined in Section \ref{sssec:th_classify} (void, nexus and
        knot) for a given eigenvalue threshold $\lambda_{\rm th}$.
    \end{enumerate}
    \begin{figure}
      \centering
      \includegraphics[width=0.49\textwidth]{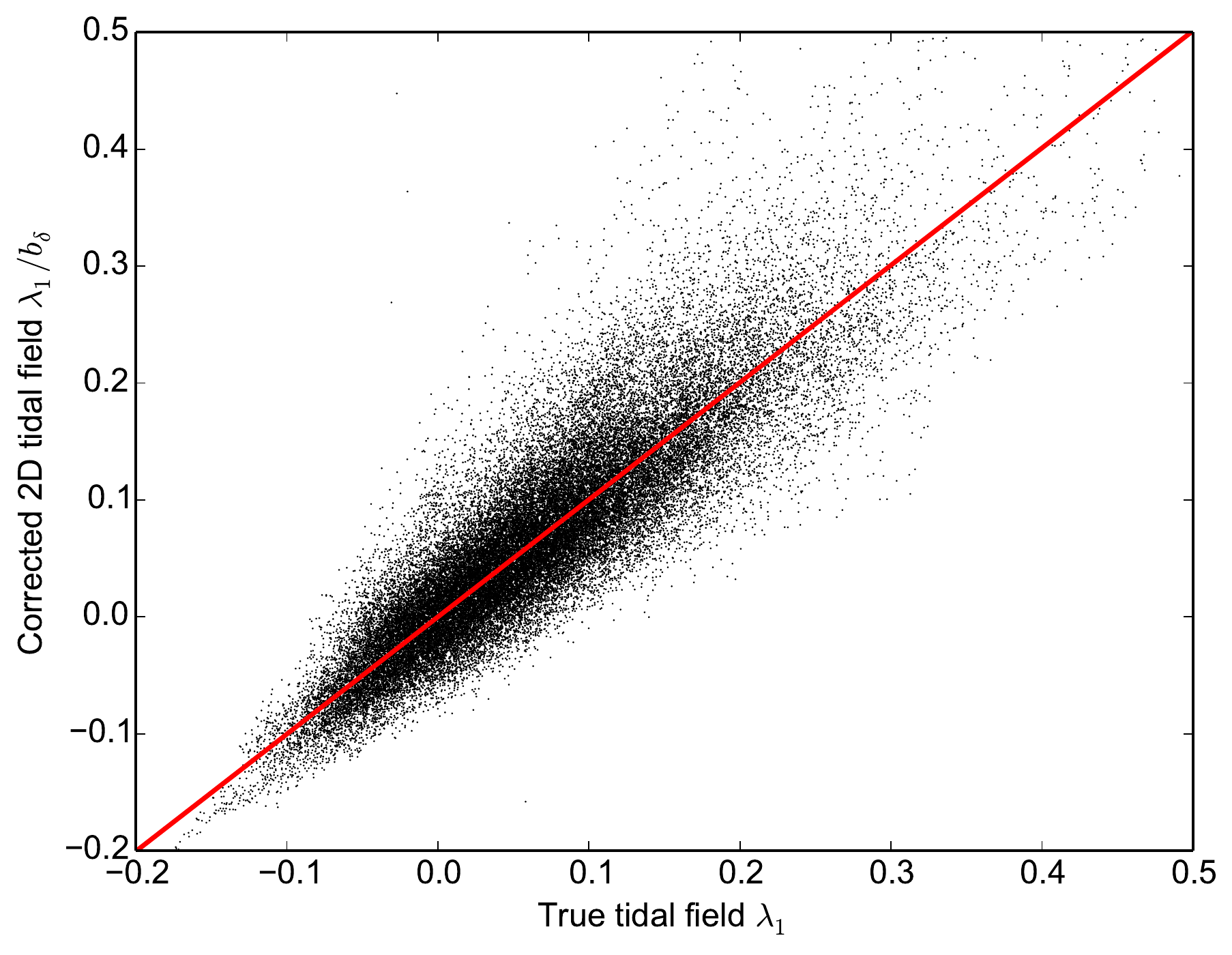}
      \caption{Two-dimensional distribution of the values of the principal eigenvalue
        $\lambda_1$ of the true projected tidal tensor (x-axis) and the 2D tidal tensor
        (y-axis) scaled by the density bias $b_\delta$. $b_\delta$ was found by fitting
        the analogous distribution in the plane of projected matter overdensity
        ($\delta_M$) and projected galaxy overdensity ($\delta_g$) to the model
        $\delta_g=b_\delta\,\delta_M$. The red solid line shows the best-fit linear
        regression, which corresponds to a slope of $1.003\pm0.002$, with a correlation
        coefficient $r\sim0.9$, similar to the correlation coefficient found in the plane
        $\delta_M-\delta_g$.}
      \label{fig:l1_fit}
    \end{figure}  
    Figure \ref{fig:sim_dens} shows the main products of this process. The top left
    panel shows a full-sky map of the density field smoothed with a $1^\circ$ kernel
    used to infer the 2D tidal tensor. The environment classification derived from
    this tidal tensor, using the fiducial eigenvalue threshold quoted in Section
    \ref{ssec:2m_results} is shown in the top right panel. Finally, the two bottom
    panels show full-sky maps of the two eigenvalues of the 2D tidal tensor.  

    For this dataset, Figure \ref{fig:sim_gauss} shows the fraction of the sky classified
    as each of the three environments as a function of the eigenvalue threshold
    $\lambda_{\rm th}$ for $1^\circ$ (left panel) and $5^\circ$ (right panel) Gaussian
    smoothing kernels. The figure also shows the prediction from Gaussian theory outlined
    in Appendix \ref{app:th_gauss} in both cases. The $1^\circ$-smoothed density field is
    clearly non-Gaussian and therefore the sky fractions are only in rough agreement
    with the Gaussian theory. After filtering out the smallest scales, responsible for most
    of the non-Gaussianity, we find that the Gaussian theory is able to describe the
    statistics of the projected cosmic web reasonably well for the $5^\circ$-smoothed field.

  \subsection{Connection with the projected tidal field}\label{ssec:2d23dsim}
    By using the true matter distribution available in the N-body simulation used to generate
    the mock galaxy catalog we can test in practice the relation between the 2D tidal field
    and the true projected tidal forces as described in Section \ref{sssec:th_2d_phys}. In order
    to do this we have carried out the following exercise:
    \begin{itemize}
     \item We first construct the three-dimensional density field of the simulation by
           interpolating the matter particles onto a Cartesian grid using cloud-in-cell
           interpolation. This was done using a grid with 512 grid points per dimension,
           corresponding to a resolution $\Delta x=1.37\,{\rm Mpc}/h$. We replicate this
           density grid in the three dimensions as described in section \ref{ssec:sim_description}
           to cover the volume of 2MASS.
     \item At each grid point we compute the value of the 3D tidal field by inverting Poisson's
           equation. Then, placing the observer at the centre of the simulated volume, we compute
           the transverse (angular) components of the 3D tidal field, as defined by the observer,
           by performing the corresponding three-dimensional rotation.
     \item We define a number $N_z$ of angular pixel maps at different redshifts sampling the
           volume covered by 2MASS. For this we used $N_z=1024$ maps uniformly distributed in the
           range $z\in[0,0.3]$, each of them with an angular resolution {\tt HEALPix} parameter
           $N_{\rm side}=1024$. The values of $T_{ab}$ computed in the Cartesian grid are then
           interpolated onto these spherical maps using trilinear interpolation. The high angular
           and radial resolution of the maps ($\Delta x_\parallel\simeq0.82\,{\rm Mpc}/h$,
           $\Delta x_\perp<0.8\,{\rm Mpc}/h$) guarantees that essentially no information is lost in the
           process.
     \item The projected tidal tensor is then estimated by performing a weighted average
           over the $N_z$ maps with weights corresponding to the survey redshift distribution:
           \begin{equation}
             \hat{\tilde{t}}_p=\frac{\sum_{i=1}^{N_z}n(z_i)\,\hat{T}_p^{(i)}}
             {\sum_{i=1}^{N_z}n(z_i)},
           \end{equation}
           where $\hat{T}_p^{(i)}$ are the transverse components of the 3D tidal tensor
           in the $i$-th pixel map at pixel $p$, and $n(z_i)$ is the number of galaxies found
           in the $i$-th redshift bin.
    \end{itemize}
    The projected tidal field thus computed can then be compared with the 2D tidal tensor
    estimated from the mock galaxy catalog as described in section \ref{ssec:sim_ideal}. Figure
    \ref{fig:eig_2dvs3d} shows the full-sky maps of the largest eigenvalue of both tensors for
    a smoothing scale $\theta_{\rm sm}=2^\circ$. The result supports the physical interpretation
    described in Section \ref{sssec:th_2d_phys}: at most scales of interest the 2D tidal tensor
    is a biased representation of the projected tidal forces, with the bias factor corresponding
    to the bias of the galaxy sample used to compute it.
    
    In order to prove this quantitatively we have carried out the following exercise: we first
    compare the values of the projected galaxy overdensity field $\delta_g$ and the true projected
    matter overdensity $\delta_M$, computed by averaging the three-dimensional matter overdensity
    in the simulation along the line of sight. From these data we estimate the density bias
    $b_\delta$ by fitting a linear model $\delta_g=b_\delta\,\delta_M$, finding a best-fit value
    $b_\delta=1.15$ with a correlation coefficient $r=0.9$. We then rescale the principal
    eigenvalue of the 2D tidal tensor in the simulation by $b_\delta$ and compare the result with
    the principal eigenvalue of the true projected tidal tensor computed as described above. The
    result is shown in Figure \ref{fig:l1_fit}, where the black dots correspond to the pairs of
    values found in the simulation, and the solid red line is the best-fit zero-intercept linear
    regression of the points. This fit yields a slope $1.003\pm0.002$, compatible with 1, and a
    correlation coefficient $r=0.88$, similar to the value found for the overdensity field.

  \subsection{Dealing with an incomplete sky coverage}\label{ssec:sim_mask}
    Even though the 2MASS catalog covered the whole celestial sphere, the Milky way
    covers a significant fraction of it, through which the density of detected sources
    is severely biased by star obscuration and dust extinction. These areas, as well as any
    region dominated by other observational systematics must therefore be discarded from the
    analysis, which complicates the application of the method presented here. The main dificulty
    lies in computing the 2D potential and its derivatives in an incomplete sky: as explained in
    the previous section, both operations are performed in harmonic space, which involves computing
    harmonic coefficients of incomplete maps that could be potentially biased. Even solving
    both problems in real space (e.g. solving Poisson's equation using relaxation techniques)
    would require assuming something about the values of the density field in the masked pixels,
    which could catastrophically bias the estimate of the 2D tidal tensor.

    In this work we have studied two different methods to deal with these issues, which we
    describe here:
    \begin{description}
      \item[Method I:] The overdensity field that we smooth and then use to compute the 2D
        potential is simply the masked overdensity field, with all masked pixels set to zero. 
      \item[Method II:] In this case we try to make use of constrained 
        Gaussian realizations (CR from here on) in order to infer the most likely value of
        the density field in the masked pixels based on the information we have about it
        outside the mask. Gaussian constrained realizations are used routinely in CMB 
        experiments to simplify the computation of the angular power spectrum of maps with
        small masked areas in them. We will outline the basic procedure used for generating
        them here, and the reader is referred to \cite{2004ApJS..155..227E} for further
        details.
        
        Writing the full-sky map of the observed density field as a vector ${\bf d}$ with
        $n_{\rm pix}$ elements, we can separate it into uncorrelated signal and noise
        components, ${\bf d}={\bf s}+{\bf n}$, where unseen (masked) pixels can be modelled
        as having a very large (infinite) noise component. Assuming both ${\bf s}$ and
        ${\bf n}$ to be Gaussianly distributed, it is easy to prove that the posterior
        probability distribution for the signal given the data is given by a multivariate
        normal distribution
        \begin{equation}
          p({\bf s}|{\bf d})=N({\bf m},\hat{C}),
        \end{equation}
        with mean and covariance given by:
        \begin{align}\label{eq:mean_cr}
          &{\bf m}=(\hat{S}^{-1}+\hat{N}^{-1})^{-1} \hat{N}^{-1}\,{\bf d}\\
          &\hat{C}=(\hat{S}^{-1}+\hat{N}^{-1})^{-1},
        \end{align}
        where $\hat{S}\equiv\langle{\bf s}\,{\bf s}^T\rangle$
        and $\hat{N}\equiv\langle{\bf n}\,{\bf n}^T\rangle$ are the signal and noise covariance
        matrices (note that the mean ${\bf m}$ corresponds to a Wiener-filtered version of the
        data). In our case we assume that the noise is white (i.e. $\hat{N}$ is diagonal)
        and we mimic the effect of the mask by making the noise variance infinite in the masked
        pixels. The inverse noise matrix $\hat{N}^{-1}$ is then easy to calculate in real space
        and is equal to the inverse noise variance in unmasked pixels and 0 in the masked ones.
        On the other hand the signal covariance $\hat{S}$ is given by the two-point correlation
        function, which makes $\hat{S}^{-1}$ easy to calculate in harmonic space, where it is
        diagonal. We can then find a maximum likelihood estimator (MLE) for the
        signal given the data as the mean of the probability distribution above (Eq.
        \ref{eq:mean_cr}). Thus, the combined inverse covariance $(\hat{S}^{-1}+
        \hat{N}^{-1})^{-1}$ needed to compute ${\bf m}$ is not diagonal in either real or
        harmonic space, and the MLE estimator must be computed by numerically solving the
        linear system:
        \begin{equation}
          (\hat{S}^{-1}+\hat{N}^{-1})\,{\bf m}=\hat{N}^{-1}\,{\bf d}.
        \end{equation}
        For this we use the conjugate gradients method, using $\hat{S}$ as a preconditioner.
        
        Finally, we must note that the action of the signal inverse covariance $\hat{S}^{-1}$ was
        computed by multiplying by the inverse power spectrum of the data in harmonic space.
        This was estimated as a polynomial fit in logarithmic space to the angular power
        specrum of the data computed from the masked overdensity field as
        \begin{equation}
          C_\ell=\frac{\sum_{m=-\ell}^\ell |\tilde{a}_{\ell m}|^2}{(2\ell+1)\,f_{\rm sky}},
        \end{equation}
        where $\tilde{a}_{\ell m}$ are the harmonic coefficients of the masked overdensity
        field (i.e. with all masked pixels set to zero) and $f_{\rm sky}$ is the fraction of
        unmasked sky\footnote{Although this is, in general known to be a biased estimator of
        the angular power spectrum for incomplete skies, we verified that the resulting power
        spectrum was in good agreement with the one estimated using more sophisticated
        algorithms (e.g. \cite{2004MNRAS.350..914C}).}.

        The assumption that the signal, noise and data are Gaussianly distributed is not correct
        for the projected overdensity field, especially at low redshifts. In order to ameliorate
        this problem, we first transformed the original overdensity field ($\delta_{\rm data}$)
        into a ``Gaussianized'' version of it given by
        \begin{equation}\label{eq:lnm1}
          \delta_{\rm Gaussian} = \ln \left[(1+\delta_{\rm data})
                                  \sqrt{1+\sigma_\delta^2}\right],
        \end{equation}
        where $\sigma_\delta^2\equiv\langle\delta_{\rm data}^2\rangle$. Since the overdensity
        field is known to be qualitatively well described by a lognormal distribution
        \cite{1991MNRAS.248....1C} (at least at the one-point level), the idea behind this
        operation is to produce a more Gaussian field by performing an inverse lognormal
        transformation on the original one. $\delta_{\rm Gaussian}$ and its power spectrum are
        then used to generate the MLE constrained realization, which is then transformed into
        a physical overdensity by inverting Eq. \ref{eq:lnm1}.
        
        Before ending this description, it is worth noting that this procedure yields the
        maximum-likelihood estimate of the tidal field inside the mask. We can then quantify
        the errors on this estimate by sampling from the posterior distribution and computing
        the standard deviation of the samples. Each sample can be drawn by generating two
        white Gaussian random fields with unit variance, ${\bf r}_S$ and ${\bf r}_N$, and
        solving the modified linear system
        \begin{equation}\label{eq:cr_random}
          (\hat{S}^{-1}+\hat{N}^{-1})\,{\bf s}=\hat{N}^{-1}\,{\bf d}+
          \hat{S}^{-1/2}{\bf r}_S+\hat{N}^{-1/2}{\bf r}_N.
        \end{equation}
        We will use this method to compute the errors in our estimate of the 2D tidal field
        for the 2MASS galaxy survey.
      \end{description}
    \begin{figure}
      \centering
      \includegraphics[width=0.5\textwidth]{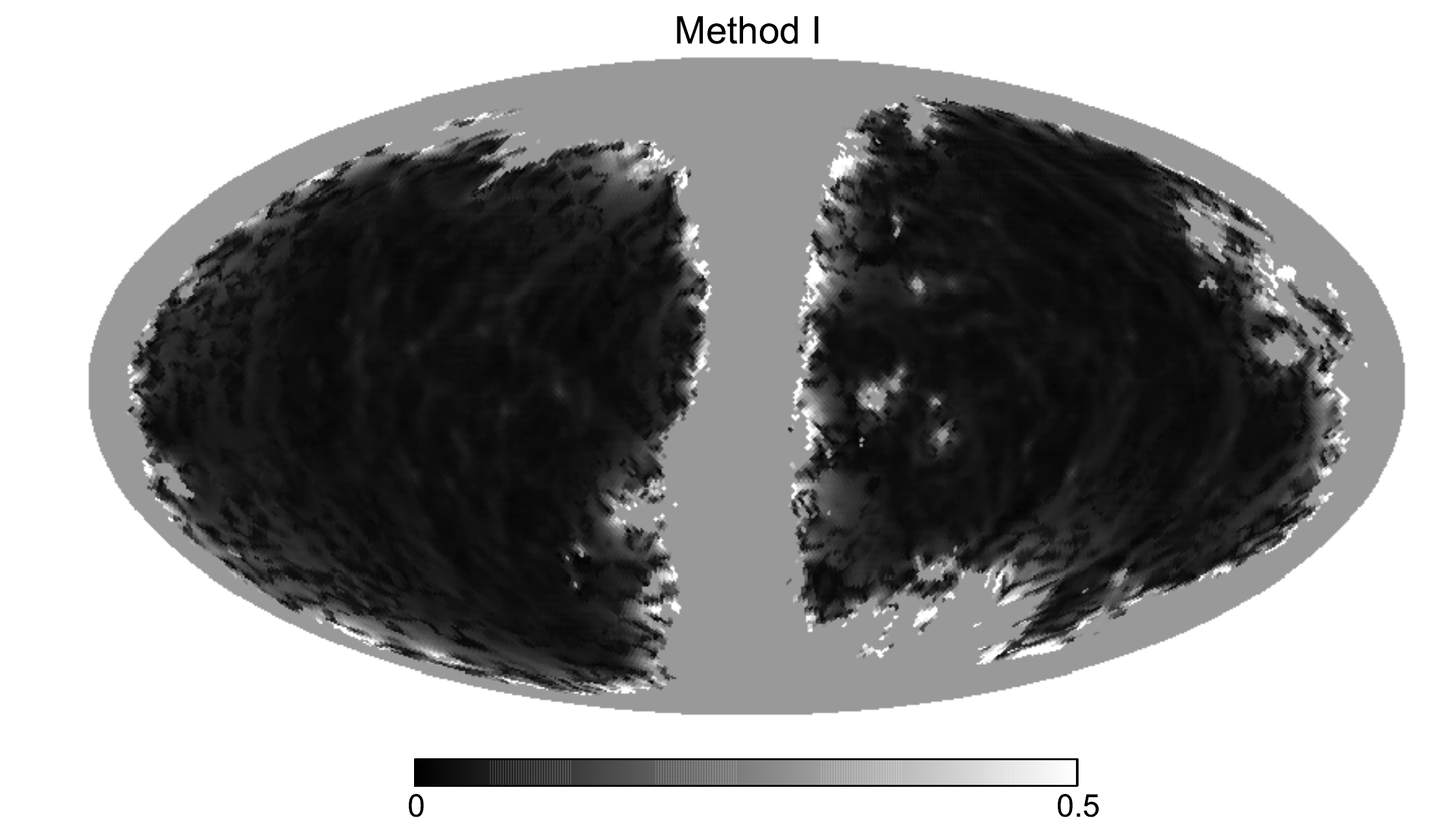}
      \includegraphics[width=0.5\textwidth]{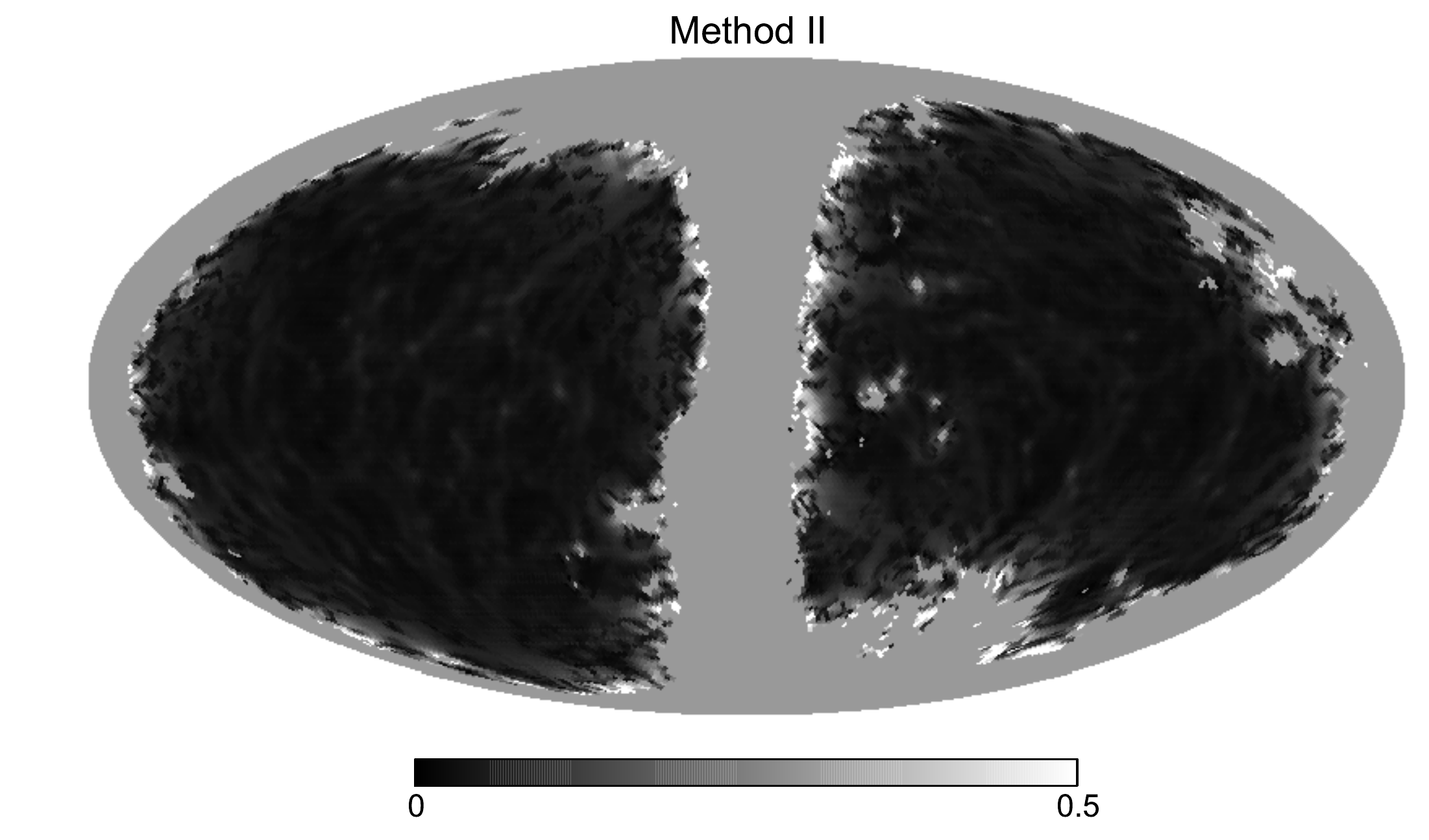}
      \caption{Maps of the error in the eigenvalues of the 2D tidal tensor in each pixel for
        the two methods described in Section \ref{ssec:sim_mask} to account for the incomplete
        sky coverage. This uses the simulations described in the text, with the low-latitude
        mask used for 2MASS, described in Section \ref{ssec:2m_describe}. Method II gives smaller
        errors overall, and we use it as our method of choice for the analysis of the 2MASS
        data.}
      \label{fig:eig_err}
    \end{figure}  
    \begin{table}
      \setlength\extrarowheight{3pt}
      \setlength{\tabcolsep}{10pt}
      \centering {
      \begin{tabular}{|c|c|c|}
        \hline
                                                   & Method I & Method II \\\hline
        $\langle\Sigma_\lambda\rangle$            & 0.074    & 0.067     \\
        \% area with $\Sigma_\lambda > 0.1$       & 16.4\%   & 13.3\%    \\
        \% area with $\Sigma_\lambda > 0.2$       & 7.0\%    & 5.9\%     \\
        \% missclassifications                    & 5.2\%    & 4.7\%     \\
        \hline
      \end{tabular}
      }
      \caption{Comparison between the two methods used to deal with the mask.
               Even though both methods show very similar results, Method II
               outperforms the other two in the four different metrics, and was
               therefore our choice in the treatment of the real data.}
      \label{tab:errors}
    \end{table}
    Both for generating constrained realizations and for the calculation of the tidal tensor
    we make extensive use of direct and inverse spherical harmonic transforms (SHTs). Unlike
    in the case of Fourier transforms, direct and inverse SHTs do not cancel each other exactly,
    and small numerical errors can be generated if many consecutive transforms are applied to a
    given map, especially towards the poles of the sphere. Although we have verified that these
    errors are sufficiently small to be almost negligible for our purposes, we have tried to
    further minimize their effect by using angular coordinates such that the masked areas
    around the galactic plane occupy the regions close to the poles of our coordinate system,
    where these errors can be most relevant. Thus, all maps displayed below were are shown in
    galactic coordinates rotated by 90 degrees (i.e., the galactic plane runs vertically
    through the centre of our maps leaving the North and South Galactic Poles to the left
    and right respectively).

    In order to evaluate the goodness of both methods we have applied them to our
    simulated catalog using the angular mask employed in the analysis of the 2MASS data
    (see Section \ref{ssec:2m_describe}). We then compare the recovered 2D tidal tensor
    with the true tidal tensor computed without the angular mask. For each method
    we compute a map containing, for each pixel, the relative error in the estimated tidal
    tensor eigenvalues, defined as
    \begin{equation}
      \Sigma_\lambda=\sqrt{\frac{(\lambda_1^t-\lambda_1^r)^2+(\lambda_2^t-\lambda_2^r)^2}
      {\langle(\lambda_1^t)^2+(\lambda_1^t)^2\rangle}},
    \end{equation}
    where $\lambda_i^t$ is the true $i-$th eigenvalue (i.e. computed from the unmasked density
    field), and $\lambda_i^r$ is the one recovered from the masked map. The ensemble average in
    the denominator was computed by summing over all pixels in the true map. We then judged the
    performance of each method by comparing four quantities: the average $\Sigma_\lambda$ across
    the sky, the fraction of the sky where $\Sigma_\lambda>0.2$ and $\Sigma_\lambda>0.1$ and the
    fraction of pixels for which the environment type differs from the one found for the true map.
    We carried out this exercise for our fiducial smoothing scale and eigenvalue threshold
    ($1^\circ$ and $\lambda_{\rm th}=0.05$).
    
    Figure \ref{fig:eig_err} shows maps of $\Sigma_\lambda$ for the three methods, and the
    quantitative results are summarized in Table \ref{tab:errors}. Overall the best performance
    is achieved by method II. We observe a mild improvement with respect to method I due to
    the ability of the maximum-likelihood constrained realization to infer the value of the
    density field in the pixels near the edge of the mask, and therefore we used that method in
    the analysis of the 2MASS data.

\section{The cosmic web of 2MASS}\label{sec:2mass}
  \begin{figure*}
    \centering
    \includegraphics[width=0.49\textwidth]{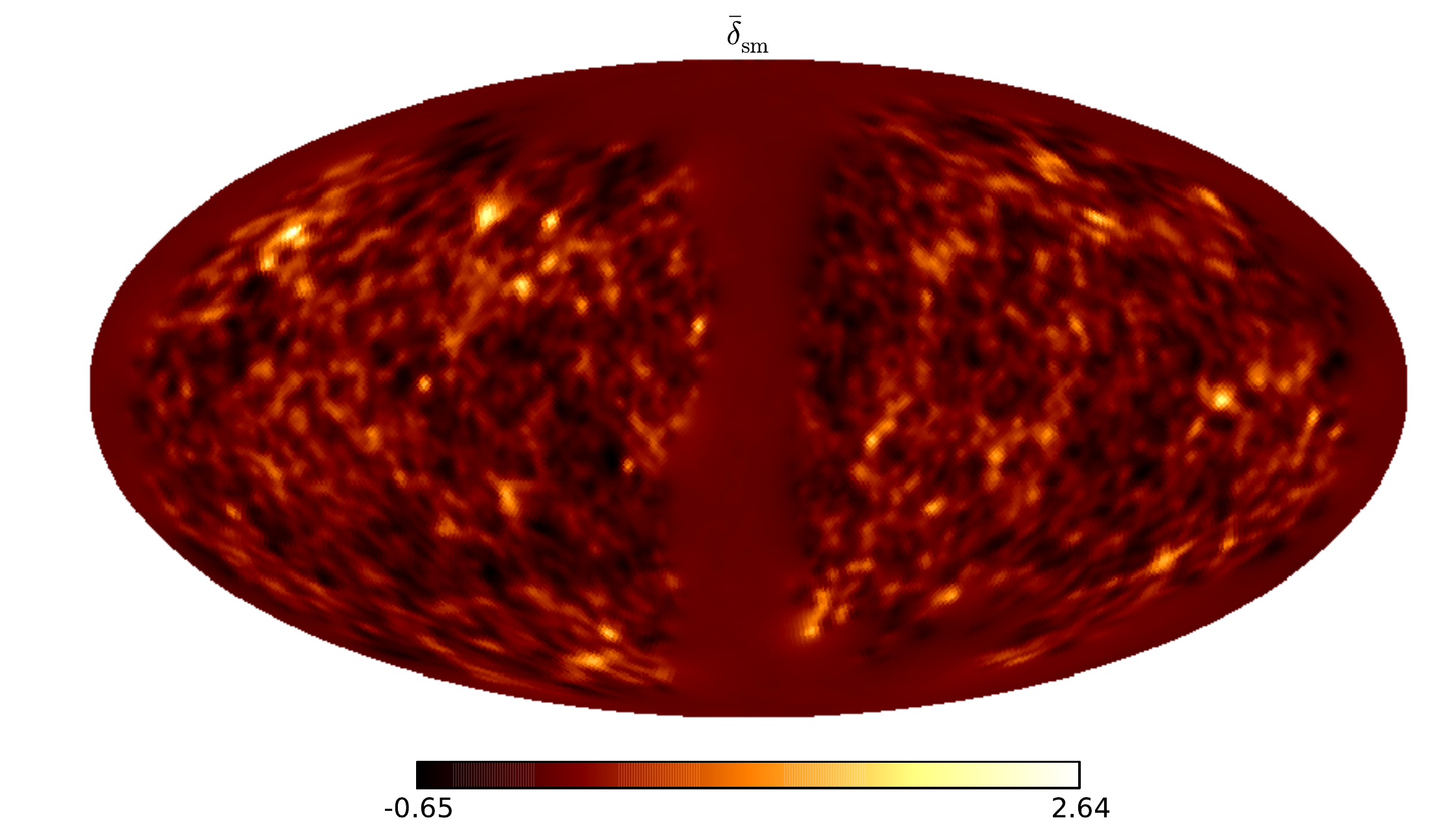}
    \includegraphics[width=0.49\textwidth]{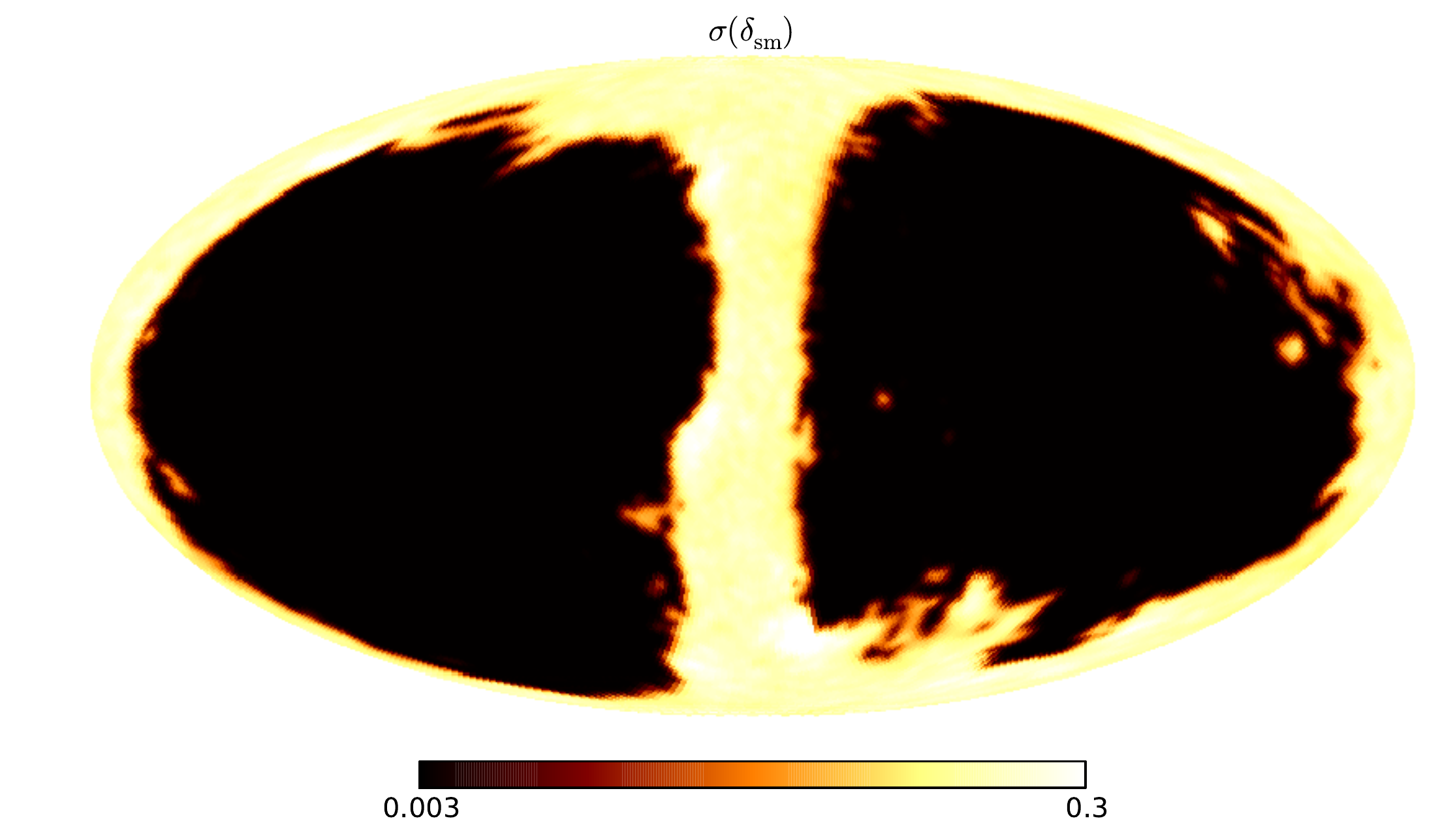}
    \includegraphics[width=0.49\textwidth]{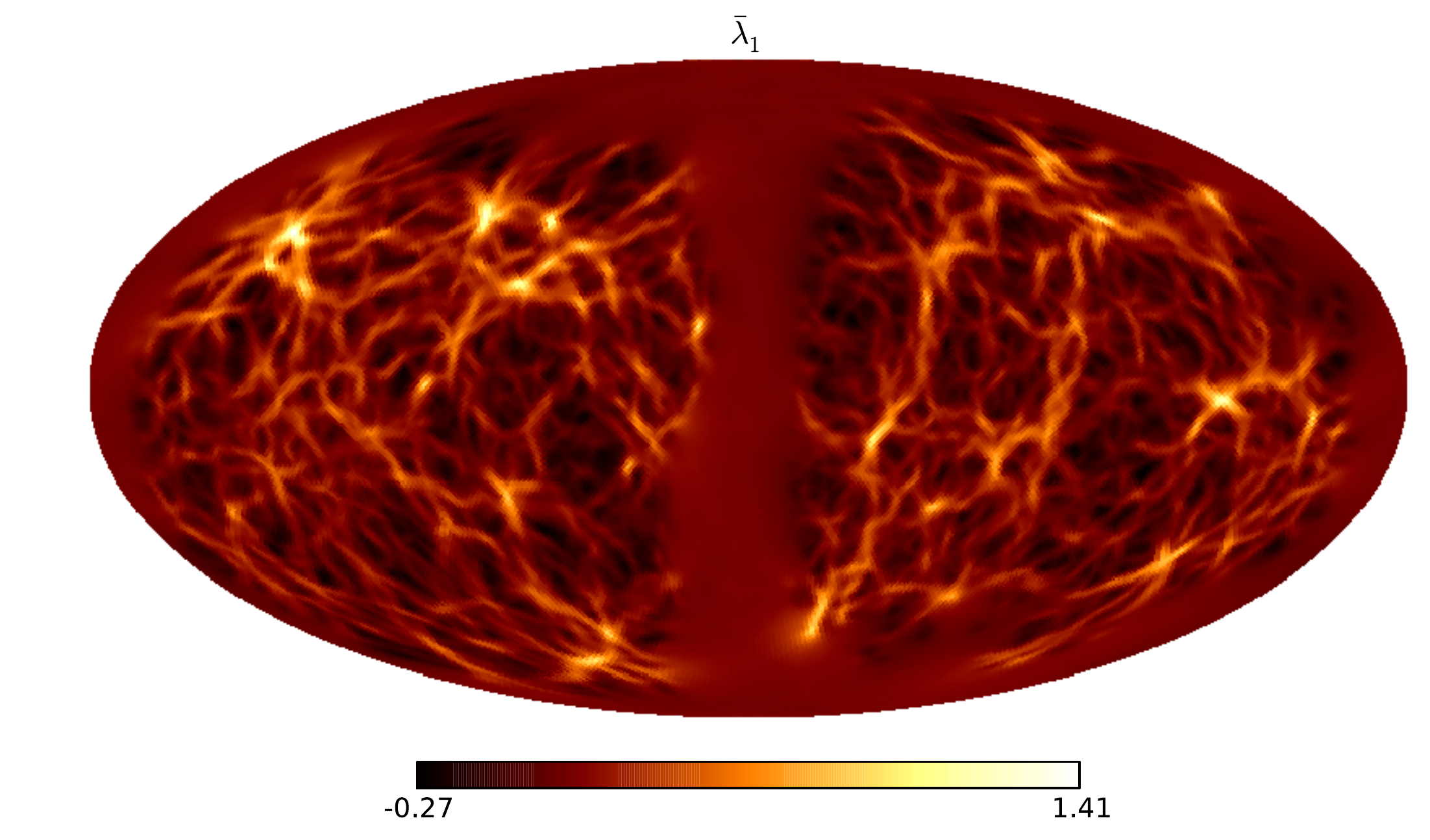}
    \includegraphics[width=0.49\textwidth]{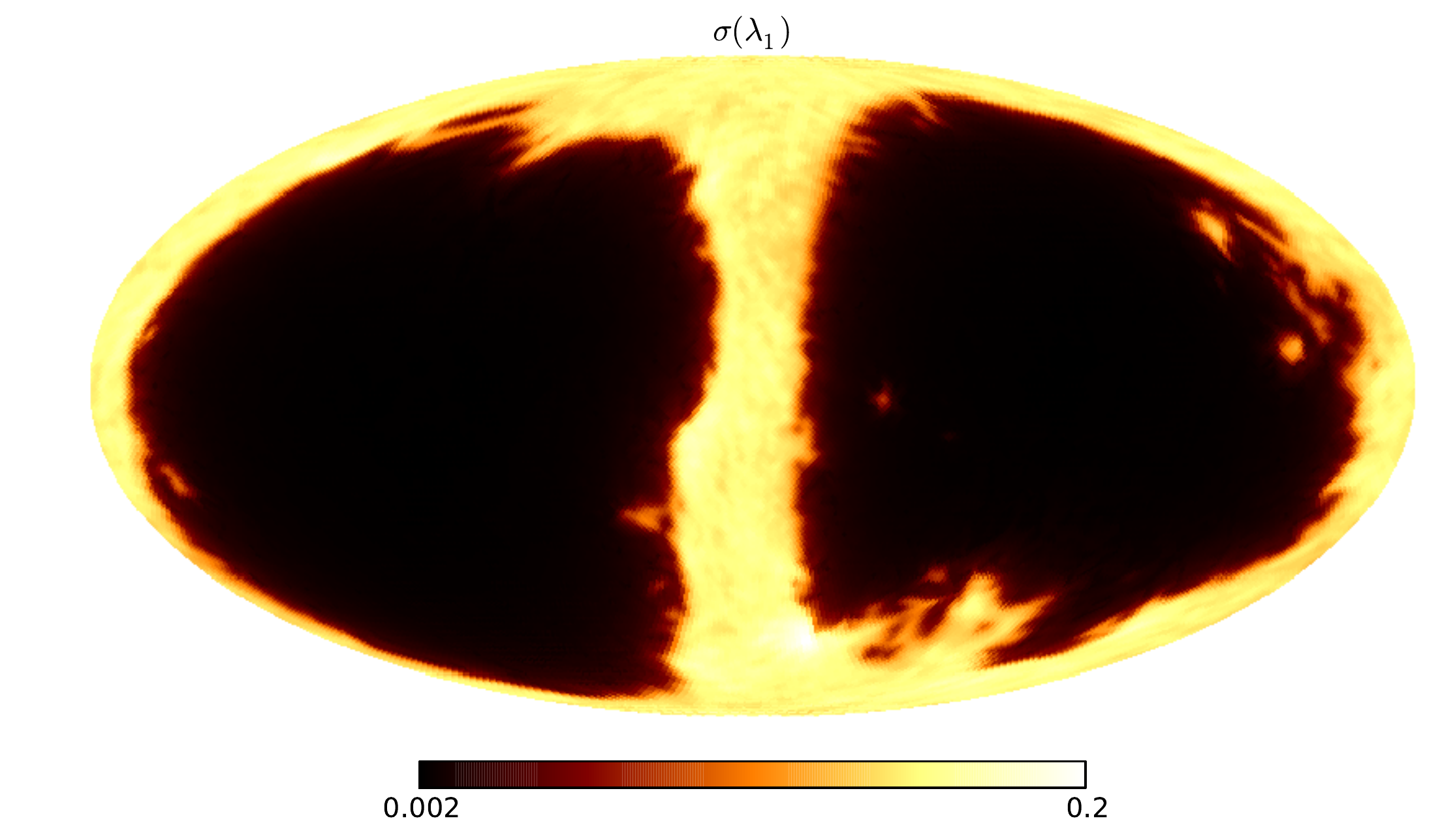}
    \includegraphics[width=0.49\textwidth]{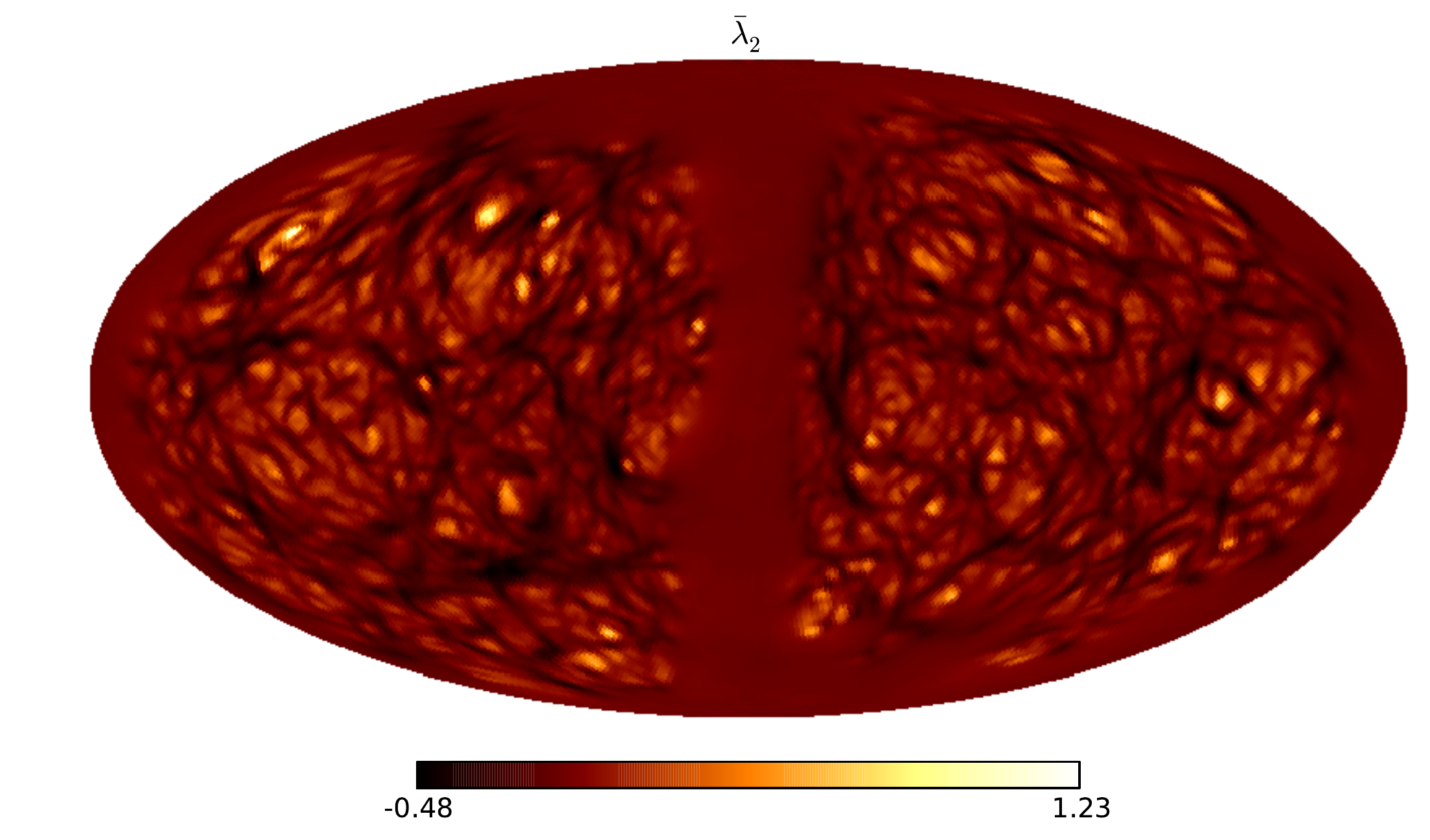}
    \includegraphics[width=0.49\textwidth]{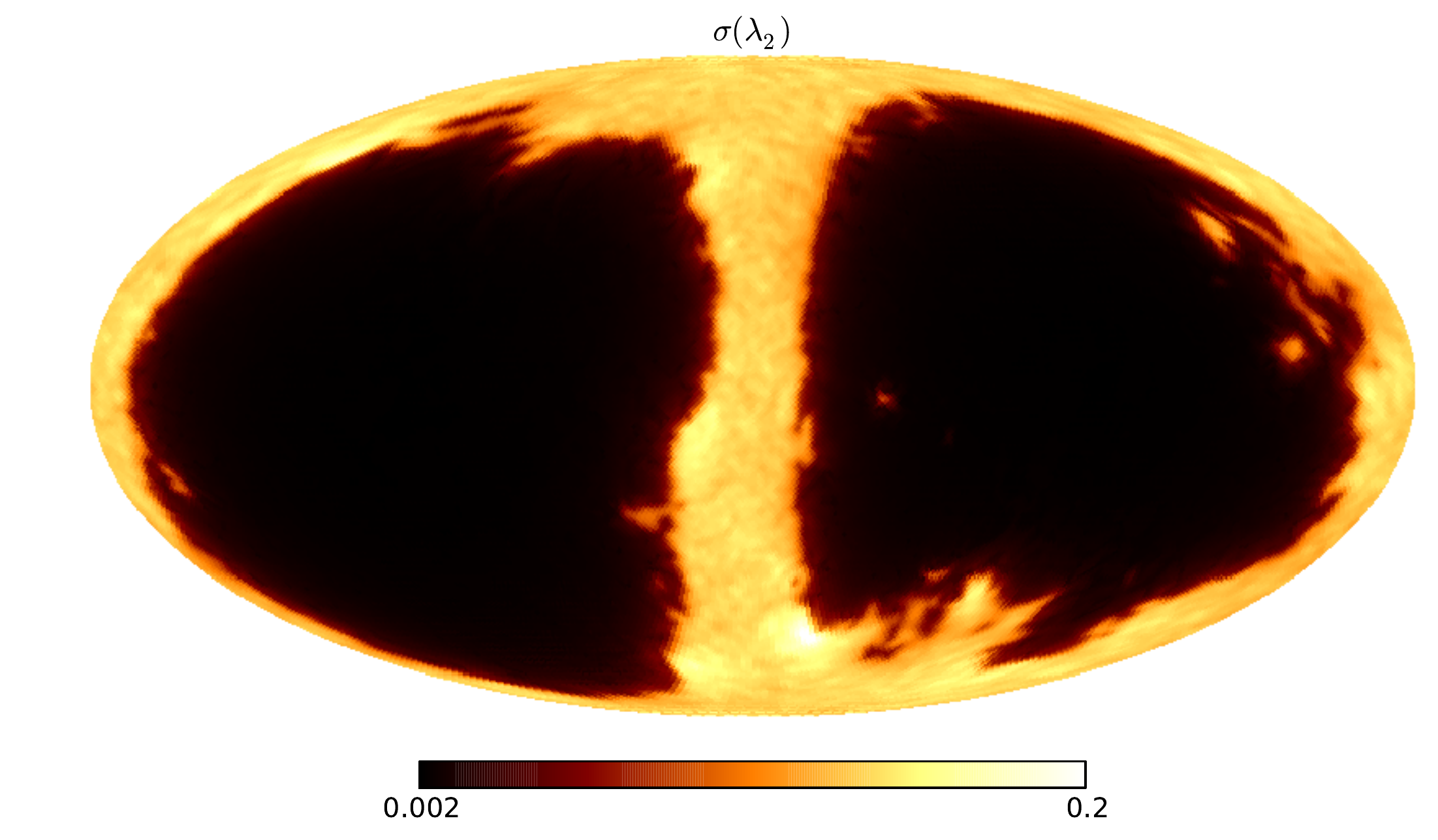}
    \caption{Mean (left column) and standard deviation (right column) of the projected
             smoothed galaxy density (upper panels) and the corresponding eigenvalues of
             the 2D tidal tensor (middle and lower panels) as estimated from 1000
             constrained realizations of the 2MASS galaxy distribution. Note that
             the density field is basically unconstrained on masked regions distant from
             the mask edges, and therefore the mean of the constrained realizations is
             close to 0 there. A Gaussian smoothing kernel with standard deviation
             $\theta_s=1^\circ$ was used.}
    \label{fig:2mass_dens}
  \end{figure*}
  In this section we describe our analysis of the projected tidal structure of the 2MASS
  survey, which we use to produce a map of the projected tidal forces in the local Universe.
  As an application of these results, we further study the environmental dependence of
  the galaxy luminosity function.
  
  \subsection{The 2MASS galaxy survey}\label{ssec:2m_describe}
    The Two-Micron All Sky Survey (2MASS) is a ground-based survey that was carried out 
    between 1997 and 2001 using two twin telescopes located at Mount Hopkins, Arizona and
    Cerro Tololo, in Chile. It imaged practically the full celestial sphere in the three
    photometric NIR bands $J$, $H$ and $K_s$\footnote{All quoted apparent magnitudes correspond
    to the Vega magnitude system}. The galaxy catalog used for our analysis is
    based on the 2MASS extended source catalog (XSC), containing 1647599 sources, of which
    more than 98\% are galaxies, the remaining 2\% being mainly galactic diffuse objects.
    Of these sources, we omitted all visually confirmed non-extended Galactic sources (flag
    ${\tt vc=2}$), artifacts (${\tt cc\_flag=a,z}$), duplicates (${\tt use\_src}\neq{\tt 1}$)
    and all objects with erroneous or excessively high $J$, $H$ or $K_s$ magnitudes, resulting
    in a sample containing 1428756 objects\footnote{Note that this procedure very closely follows
    the method used by \citep{2014ApJS..210....9B} to produce the 2MASS photometric redshift
    survey.}. In the analysis of this sample we used the $K_s$-band $20\,{\rm mag}/{\rm arcsec}^2$
    isophotal fiducial elliptical aperture magnitude (${\tt k\_m\_k20fe}$), which was corrected
    for Galactic extinction as
    \begin{equation}
      K_s\rightarrow K_s-A_K,
    \end{equation}
    where the $K_s$-band correction $A_K=0.367\,E(B-V)$ was computed from the dust
    reddening maps of \cite{1998ApJ...500..525S}.
        
    The main galaxy sample used in this analysis was selected with the aim of
    obtaining a complete and homogeneous sample. To that extent we followed the
    same procedures that were used in \cite{2004PhRvD..69h3524A,2010MNRAS.406....2F,
    2015MNRAS.449..670A}, which we summarize here. The two main sources of systematic
    incompleteness are dust extinction (quantified in terms of $A_K$ above) and stars. We found
    that a complete and homogeneous sample can be selected for a limiting magnitude $K_s=13.9$
    by cutting out all regions of the sky with either $A_K\geq0.06$ or $\log_{10}(n_{\rm star}/
    {\rm deg}^2)\geq3.5$, where $n_{\rm star}$ is the counts of point sources brighter than
    $K_s=14$. This procedure reduces the fraction of useable sky to about 69\% and constitutes
    our main  galaxy sample, containing 746733 of the 983963 galaxies with $K_s\leq13.9$.
    
  \subsection{Results}\label{ssec:2m_results}
    \subsubsection{Estimating the 2D tidal tensor}\label{sssec:2m_est}
    \begin{figure}
      \centering
      \includegraphics[width=0.48\textwidth]{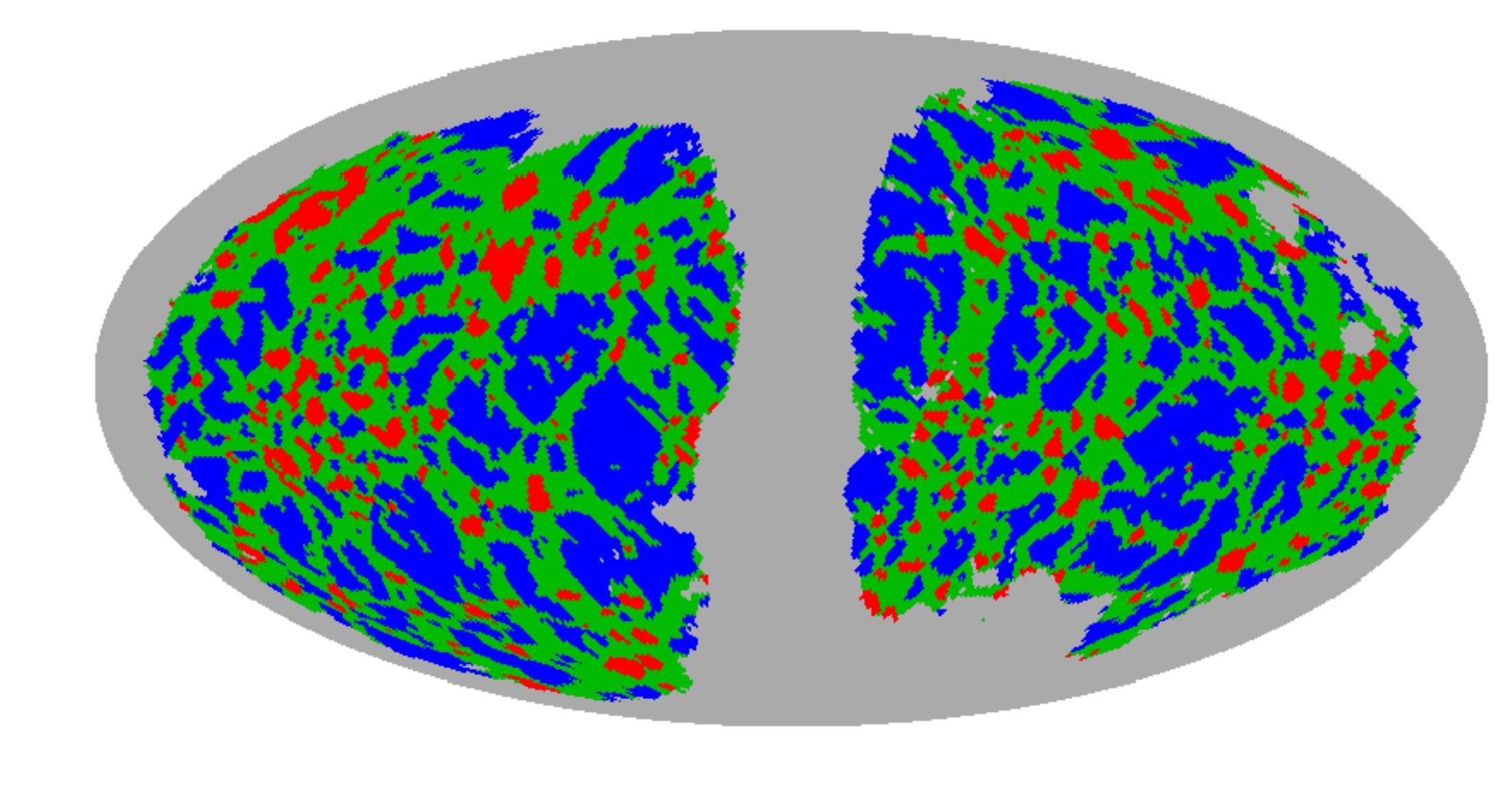}
      \caption{Tidal classification into voids (blue), nexuses (green) and knots (red) of the
               2MASS density field. This classification is based on the $1^\circ$-smoothed
               maps of the tidal field eigenvalues shown in Figure \ref{fig:2mass_dens}, with
               a threshold $\lambda_{\rm th}=0.05$.}
      \label{fig:2mass_class}
    \end{figure}
      We estimate the 2D tidal tensor for the galaxy sample described above using Method II
      outlined in Section \ref{ssec:sim_mask}. In this approach we generate a
      suite of 1000 constrained lognormal realizations of the projected galaxy density field
      compatible with its statistics in the unmasked regions. Each of these realizations is
      generated with a HEALPix resolution ${\tt nside}=64$ and further smoothed using a Gaussian
      kernel with standard deviation $\theta_{\rm sm}=1^\circ$. The smoothed density field is then
      used to estimate the 2D tidal tensor as well as its eigenvalues and eigenvector in each
      realization as outlined in Section \ref{ssec:sim_ideal}. We use this ensemble of estimated
      2D tidal fields to evaluate the uncertainty in the measurement of $\hat{t}$. Figure
      \ref{fig:2mass_dens} shows the mean (left column) and standard deviation (right column) of
      the smoothed density field (upper panels) and the major and minor eigenvalues of the 2D
      tidal tensor (middle and lower panels) computed from the constrained realizations.
      
      \begin{figure}
        \centering
        \includegraphics[width=0.49\textwidth]{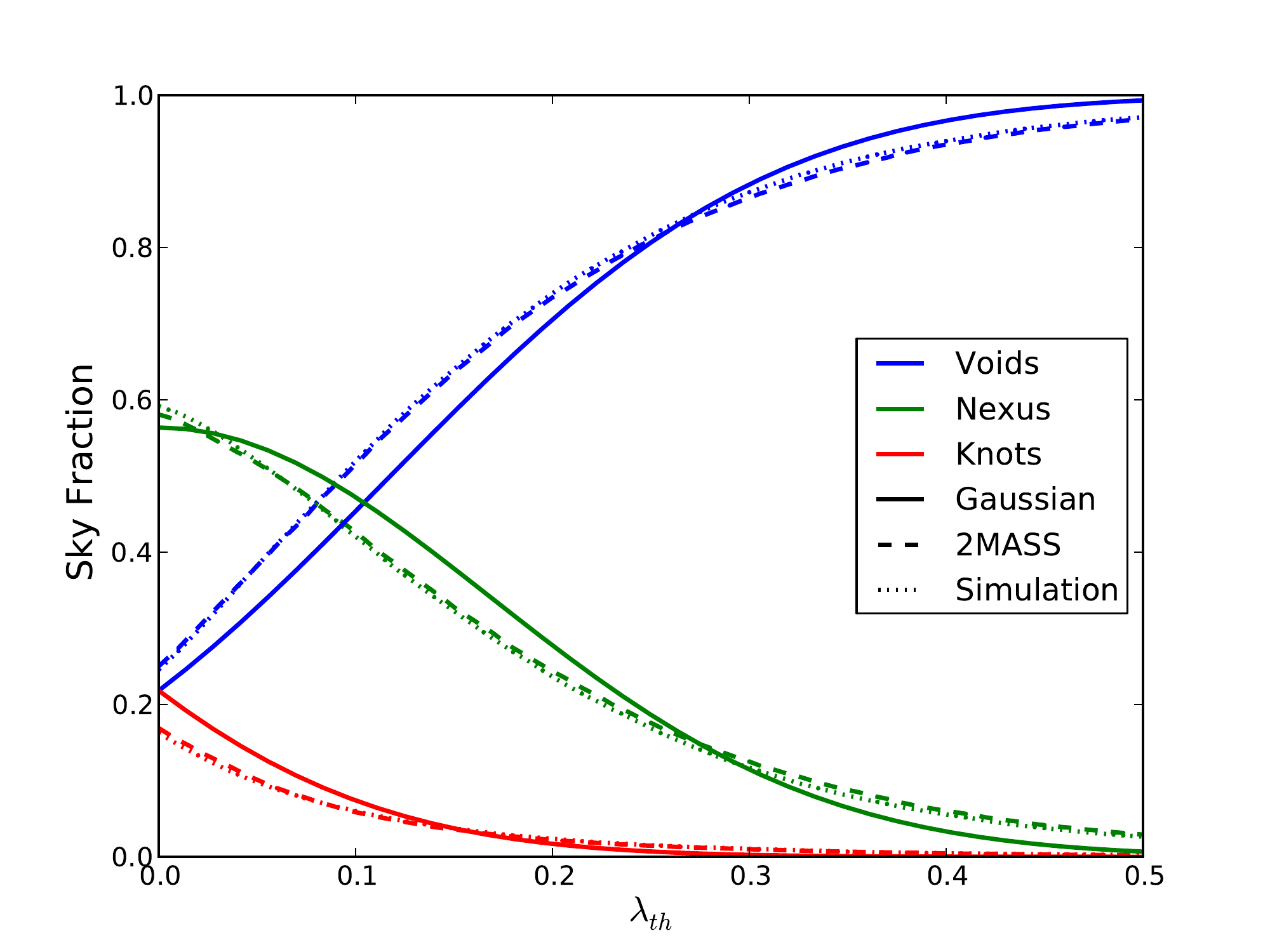}
        \includegraphics[width=0.49\textwidth]{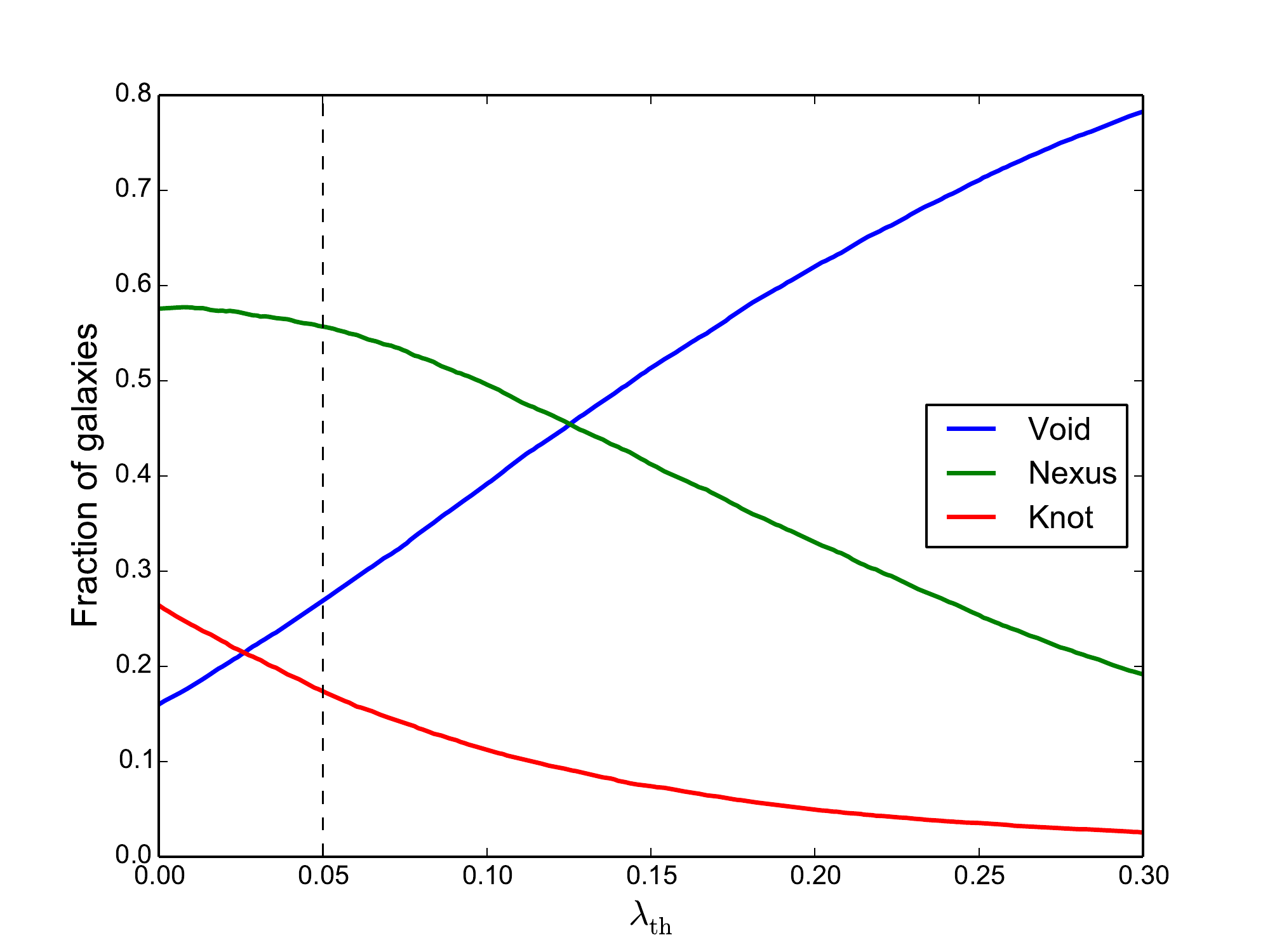}
        \caption{{\sl Top panel:} sky fraction in the different environments for the 2MASS sample 
                 (dashed lines) compared with the simulated catalog (dotted lines) and the Gaussian
                 theoretical prediction (solid lines) as a function of the eigenvalue threshold
                 and for a smoothing scale $\theta_s=1^\circ$. Blue, green and red lines correspond
                 to voids, nexuses and knots respectively. Note that the level of disagreement with
                 respect to the Gaussian theoretical prediction is similar to that shown in Fig.
                 \ref{fig:sim_gauss}. {\sl Bottom panel:} fraction of galaxies
                 in the 2MASS sample found in the three different environments as a function of
                 the eigenvalue threshold. The vertical dashed line shows our choice of
                 $\lambda_{\rm th}$.}
        \label{fig:2mass_fv}
      \end{figure}
    \begin{figure*}
      \centering
      \includegraphics[width=0.8\textwidth]{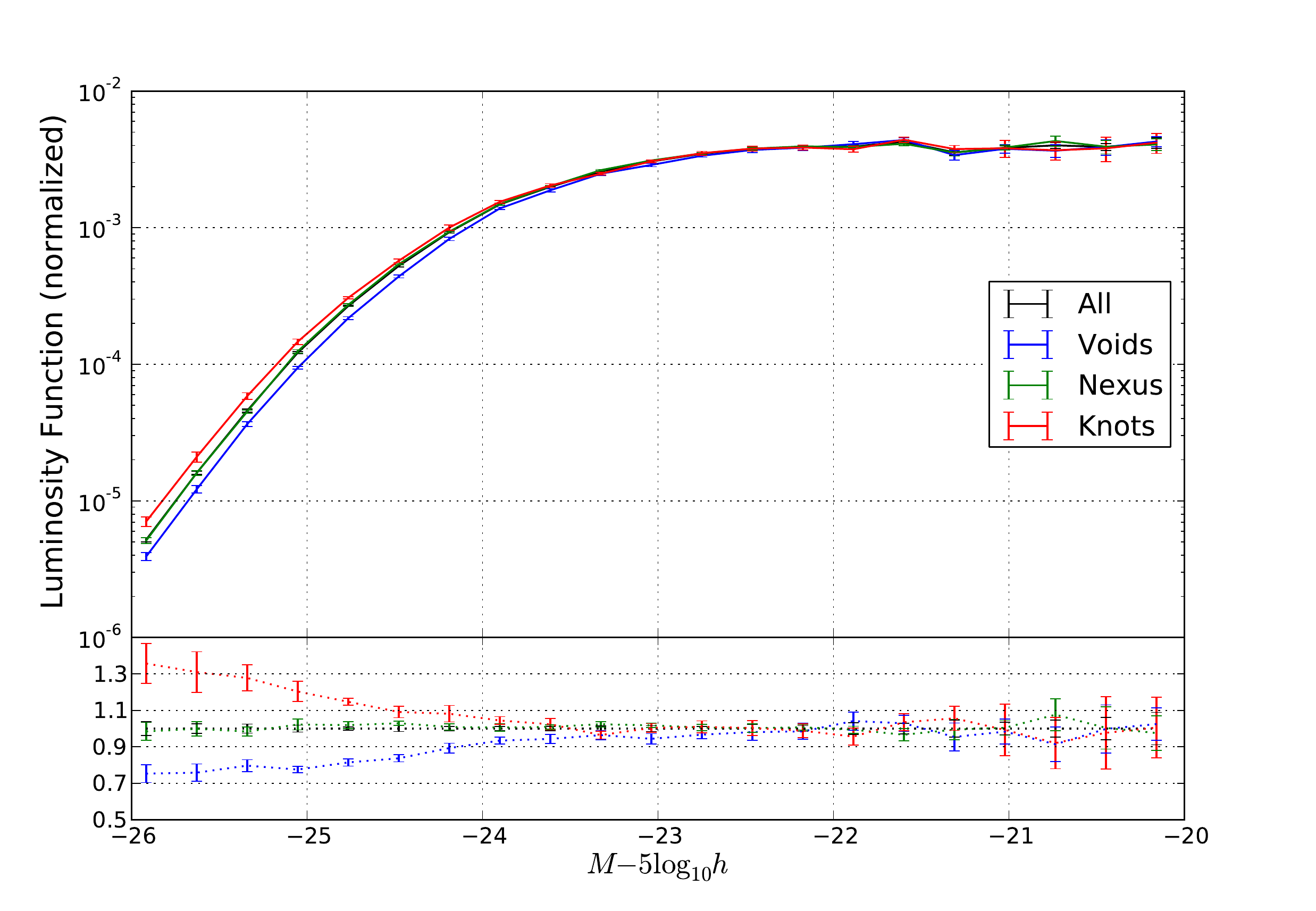}
      \caption{Luminosity function of the three environments shown in logarithmic scale. In the
               bright end we see a significant overabundance of galaxies in knots with respect
               to voids as predicted by the standard models of galaxy formation. The color code
               is voids (blue); nexuses (green); knots (red). The luminosity functions shown
               here were normalized to have the same amplitude on magnitudes $M-5\log_{10}h>-23$.
               The bottom panel shows the ratios of the void, nexus and knot LFs to the overall
               LF.} \label{fig:lumfun}
    \end{figure*}
      The uncertainty in the determination of $\hat{t}$ is very small in regions far from
      the mask edges, and grows sharply as they are approached. Knowing this
      uncertainty allows us then to define the regions where we trust that estimate enough
      for the subsequent analyses. We have thus computed the following quantity at each pixel:
      \begin{equation}
        \tilde{\Sigma}(\nv)=\sqrt{\frac{\sigma^2[\lambda_1(\nv)]+\sigma^2[\lambda_2(\nv)]}
        {\langle\bar{\lambda}_1^2+\bar{\lambda}_2^2\rangle_m}},
      \end{equation}
      where $\sigma[\lambda_i(\nv)]$ is the uncertainty in the $i$-th eigenvalue computed from
      the constrained realizations and $\langle ...\rangle_m$ implies averaging over all
      unmasked pixes. $\tilde{\Sigma}$ quantifies the magnitude of the error on the tidal
      tensor eigenvalues normalized by their typical value. In all subsequent analyses, only 
      regions not discarded by the mask described in Section \ref{ssec:2m_describe}, and for
      which $\tilde{\Sigma}<0.2$ were used, which reduced the available fraction of the sky to
      65\%. We will use this combined mask in all subsequent analyses unless otherwise stated.
      
    \subsubsection{Statistics of the cosmic web}\label{sssec:2m_cweb}
      As described in Section \ref{sssec:th_classify}, our measurement of the 2D tidal field can
      be used to define different types of environments in terms of the number of eigenvalues
      found above a given threshold $\lambda_{\rm th}$. For this we use $\lambda_{\rm th}=0.05$,
      which we determined by dividing the 2MASS galaxy sample as equally as possible between
      environments (see Section \ref{sssec:th_classify} and the bottom panel of Fig.
      \ref{fig:2mass_fv}). The environment classification thus
      found for 2MASS, using the 2D tidal tensor averaged over the 200 constrained realizations,
      is shown in Figure \ref{fig:2mass_class}.      
      
      In order to test the agreement of the statistics of the recovered tidal field with
      our theoretical expectations we have compared the sky fraction occupied  by the different
      environments as a function of $\lambda_{\rm th}$ with the results from our simulated
      galaxy catalog and the theoretical Gaussian prediction outlined in Appendix
      \ref{app:th_gauss}, as shown in Figure \ref{fig:2mass_fv}. The sky fractions 
      recovered from the data agree well with the results from the simulated catalog and,
      qualitatively, follow the same trend predicted by the Gaussian theory. However, the
      agreement with the latter is much poorer, due to the non-Gaussian nature of the density
      field on small scales.
      
    \subsubsection{Environmental Dependence of Luminosity Function}\label{sssec:2m_lfun}
    Having access to information about the tidal forces allows us, among other things, to
    study their influence in the process of galaxy formation and evolution. As an example
    of this kind of application we have studied the dependence of the luminosity function
    (LF) on the type of tidal environment. It is well known that more luminous galaxies tend to
    reside in the highest density regions of the Universe, however no clear
    dependence has yet been found on tidal or dimensional properties of the environment
    \cite{2015MNRAS.448.3665E}. In order to test this standard prediction in the context of the
    projected cosmic web, and to further search for other types of environmental dependence we
    have estimated the $K_s$-band luminosity function of 2MASS galaxies in knots, nexuses and
    voids.
    
    Given the redshift $z$ and apparent magnitude $m$ of a galaxy, its absolute magnitude
    can be computed as
    \begin{equation}\label{eq:rel2abs}
      M=m-5\,\log_{10}\left[\frac{d_L(z)}{1\,{\rm Mpc}}\right]-25+K(z)+E(z),
    \end{equation}
    where $d_L(z)\equiv(1+z)\,\chi(z)$ is the luminosity distance and $K(z)$ and $E(z)$
    are the $k$-correction and evolution correction respectively. For the $K_s$-band luminosity
    of 2MASS galaxies we use the simple forms $K(z)=-6\,\log_{10}(1+z)$
    \cite{2001ApJ...560..566K} and $E(z)=z$ \cite{2003ApJ...592..819B}, which are accurate enough
    at the low redshifts ($z\lesssim0.3$) covered by the survey. For this exercise we used a
    sample of 114930 galaxies with spectroscopic redshifts measured by SDSS
    \cite{2011AJ....142...72E} and the 2MASS redshift survey \cite{2012ApJS..199...26H}. This
    sample covers about 5000 square degrees at galactic latitudes $b\gtrsim60^\circ$ with a
    spectroscopic completeness above 90\% for $K_s<13.9$ (with the remaining 10\% not showing
    any particular bias in magnitude or position). This sample was used in the calibration of the
    2MPZ survey \cite{2014ApJS..210....9B}, and the spectroscopic redshifts used here were
    obtained from their publicly available catalog.
    
    Since the LF $\phi(M)$ is the number density of galaxies per unit interval of absolute
    magnitude $M$, the probability of finding the $i$-th galaxy with absolute magnitude $M_i$
    given its redshift $z_i$ in a magnitude-limited sample is given by
    \begin{equation}\label{eq:lf_prob}
      p(M_i|z_i)=\frac{\phi(M_i)}{\int_{-\infty}^{M_{\rm lim}(z_i)}\phi(M)dM},
    \end{equation}
    where $M_{\rm lim}(z)$ is the limiting magnitude at redshift $z$ given the magnitude
    limit of the sample. The joint likelihood of the full sample is then given by
    the product of this quantity over all the galaxies in the sample. Thus, given a
    model for the luminosity function, we can find the best-fit parameters of the model
    by maximizing $\mathcal{L}$.
    
    Although a Schechter function \cite{1976ApJ...203..297S} has often been advocated as a
    simple and accurate parametrization of the LF, we prefer to use a non-parametric model
    in order to directly study the environmental dependence as a function of luminosity.
    Thus our method of choice is the non-parametric step-wise maximum
    likelihood estimator introduced by \cite{1988MNRAS.232..431E} (EEP from here on).
    This method models the LF as a step-wise function in a number of
    magnitude bins,
    \begin{equation}
      \phi(M) = \sum_{n=1}^{N_{\rm bins}} \phi_n W(M_n-M),
    \end{equation}
    where $W(M_n-M)$ is a top-hat function centered around $M_n$ with a width $\Delta M$.
    Substituting this model in Eq. \ref{eq:lf_prob}, we find that the maximum-likelihood
    parameters $\phi_n$ must satisfy:
    \begin{equation}
      \phi_n = \frac{\sum_{m=1}^{N_{bins}}W(M_n-M_m)}{\sum_{i=1}^{N_g}
      \frac{H(M_{\rm lim}(z_i)-M_n)}{\sum_{l=1}^{N_{bins}}\phi_l H(M_{\rm lim}(z_i)-M_l)}},
    \end{equation}
    where $H(x)$ is the integral of $W(x)$. Given an initial guess for the $\phi_n$'s
    this equation can then be solved iteratively.
    
    In order to use sample with high signal-to-noise magnitudes, we only use galaxies
    brighter than $K_s=13.75$ which, after imposing the mask defined in section \ref{sssec:2m_est}
    leaves us with a sample of 92585 galaxies, 22613 of them in voids, 50427 in nexuses and 19545
    in knots. $\phi(M)$ was estimated in $N_{\rm bins}=20$ magnitude
    bins in the range $M-5\log_{10}h\in[-26,-20]$, and the statistical errors
    of these measurements were computed using 10 random jackknife realizations of each galaxy
    subsample. These jackknife errors were found to be in good agreement with the Poisson
    uncertainties expected given the galaxy counts in each magnitude bin. Since we expect a
    smaller environmental dependence in the faint end of the LF,
    we fix the normalization of the luminosity function in each environment by matching its
    amplitude to that of the overall LF for magnitudes $M-5\log_{10}h>-23$ in a $\chi^2$-sense.    
    \begin{figure}
      \centering
      \includegraphics[width=0.49\textwidth]{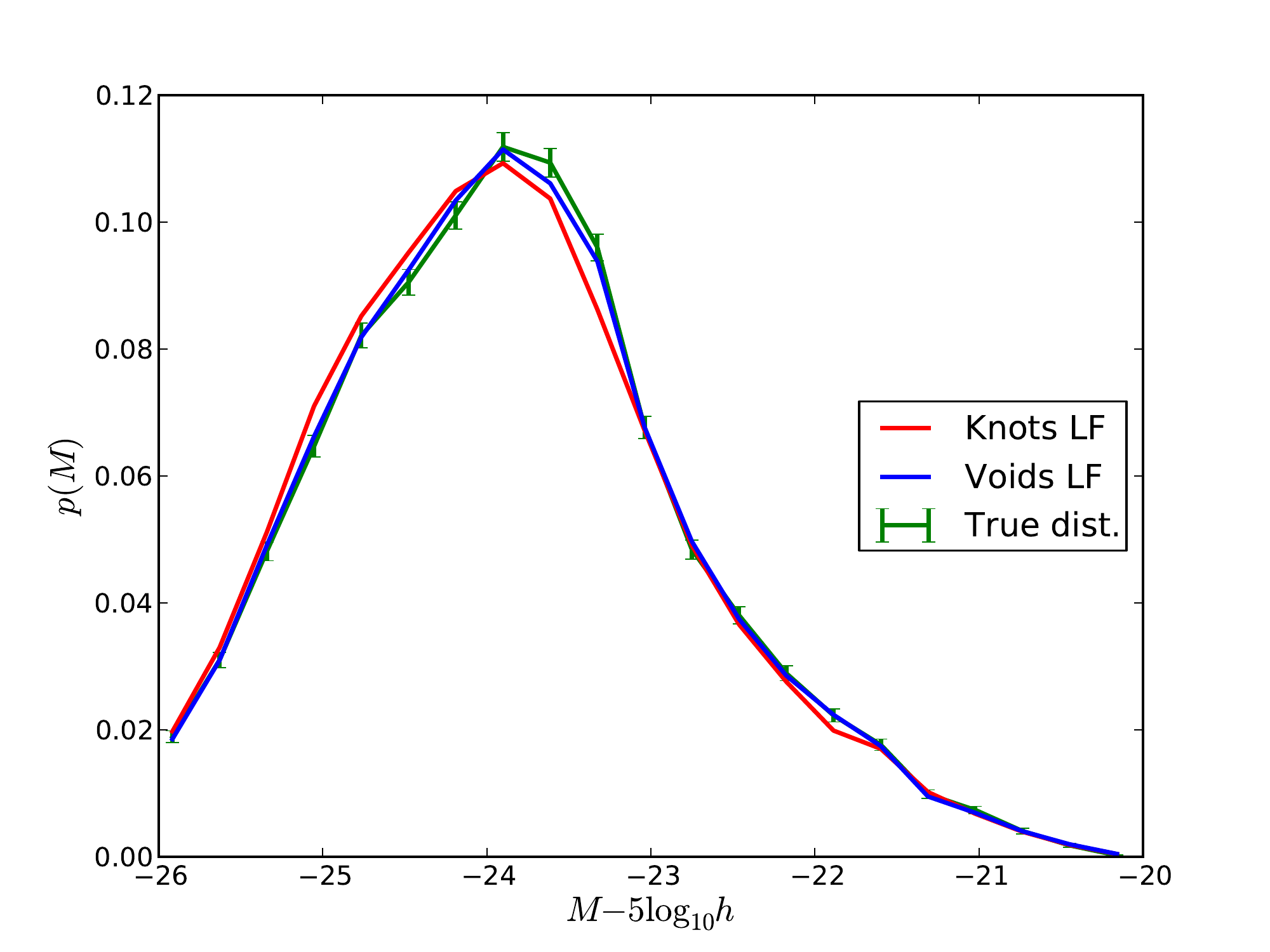}
      \caption{Magnitude distribution computed using the redshift of void galaxies and the
               EEP luminosity function estimated in voids (blue line), compared to the same
               quantity computed using the knots LF (but still using the void galaxy redshifts)
               (red line) and to the true magnitude distribution of void galaxies (green line
               with Poisson error bars). While the voids LF recovers the true
               magnitude distribution correctly, the distribution recovered using the knots
               LF shows significant disagreement. This confirms the statistical significance of
               the differences found between the luminosity function in voids and knots.}
      \label{fig:lumtest2}
    \end{figure}
    \begin{figure}
      \centering
      \includegraphics[width=0.49\textwidth]{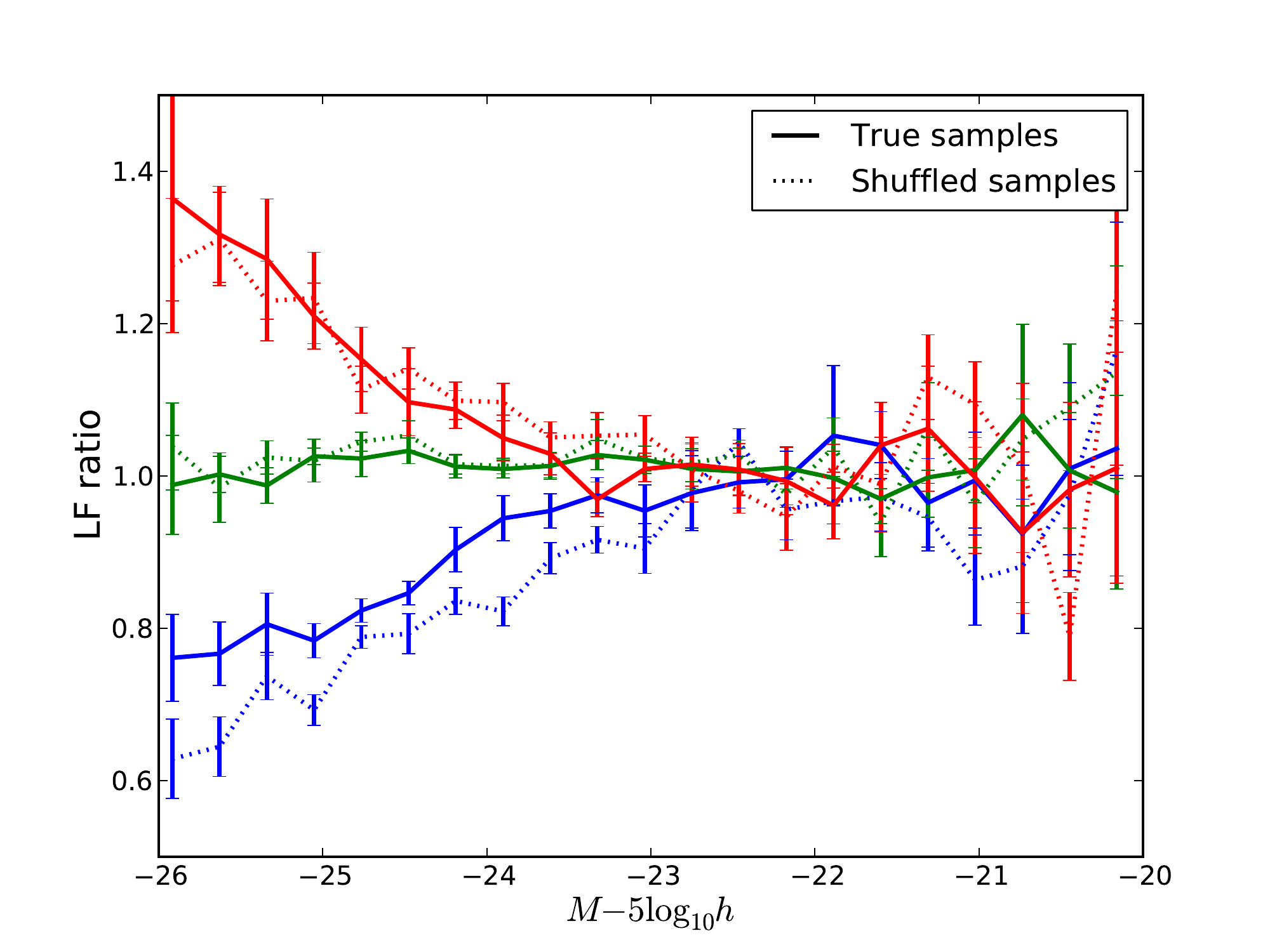}
      \caption{Ratio of the LFs for the true (solid lines) and shuffled (dotted lines) for voids
               (blue), nexuses (green) and knots (red), to the overall luminosity function. The
               true and shuffled samples show perfectly compatible luminosity functions, which
               support the idea that the differences between the three LFs is mainly due to the
               densities of the different environments, and not to their tidal structure.}
      \label{fig:lumtest}
    \end{figure}
    
    The result of this exercise is displayed in Figure \ref{fig:lumfun}: the top panel
    shows the LF measured in each environment and overall, while the bottom one shows the
    ratio of the LF in each environment with respect to the measured values across the
    whole sky. We observe a significant increase in the number density of luminous
    galaxies in knots with respect to voids, while the nexus LF in is perfectly compatible
    with the overall luminosity function across the whole range of luminosities. As a consistency
    check, and in order to verify the statistical significance of these differences we
    reconstructed the distribution of absolute magnitudes in the data from our estimates of the
    luminosity function, following \cite{1979ApJ...232..352S}. To do this, we first compute, for
    each galaxy at redshift $z_i$ in our sample, the conditional probability distribution
    $p(M|z_i)$, given in Eq. \ref{eq:lf_prob} in terms of the EEP luminosity function estimated
    for that sample. We then sum the distributions obtained for all galaxies, normalize the result
    to unity when integrated over
    magnitudes, and compare the result with the actual magnitude distribution of the data. The
    result is shown in Figure \ref{fig:lumtest2}: the true magnitude distribution for void galaxies
    is nicely matched by that assuming the voids LF. However, if the luminosity function estimated
    for knots is used instead, the recovered distribution differs significantly from the true one,
    which confirms the difference between the two luminosity functions.

    Although this is an interesting result, the fact that the three different environment
    types are associated with different density distributions (which are actually disjoint
    in the case of knots and voids), makes it difficult to ascertain whether these observed
    differences are caused by the tidal structure of the environment or merely by its local
    density, and there are theoretical and empirical reasons to presume that density is the
    dominant variable \cite{2015MNRAS.447.2683A,2015MNRAS.448.3665E}. In order to address this
    question we have carried out an analysis similar to the one used in \cite{2015MNRAS.448.3665E}.
    We start by dividing our spectroscopic sample into bins of
    density. Then we generated shuffled galaxy samples for each environment type by
    substituting each galaxy residing in that environment by a random galaxy from the density
    bin in which it belongs (which will not in general come from the same environment type). We
    then compute the LF for these shuffled samples and compare them with the corresponding
    ``true'' LF for each environment type. Any differences between both LFs would then
    be entirely due to tidal effects independent of the environmental density. The result
    is shown in Figure \ref{fig:lumtest}: in all cases the LFs for the true
    and shuffled samples are perfectly compatible. Thus there is no evidence for a dependence
    of the LF on tidal effects beyond the density dependence.

\section{Discussion}\label{sec:discussion}
  In this paper we have presented a method to reconstruct the transverse tidal forces using
  the angular position of galaxies in a survey without reliable radial information. The method is
  based simply on adapting the standard Fourier methods used to estimate the tidal field in
  three-dimensional datasets to the two-dimensional sphere. We label the object thus recovered
  the ``2D tidal tensor'' and show, both in perturbation theory and using a simulated catalog,
  that for all scales of interest it can be interpreted as being proportional to the transverse
  components of the true tidal tensor averaged along the line of sight with the survey
  window function.

  Since the method makes extensive use of operations in harmonic space, special care must be
  taken when dealing with incomplete sky coverage. In order to deal with this we make use of
  constrained lognormal realizations (as described in Section \ref{ssec:sim_mask}) in order
  to find a maximum-likelihood estimate of the tidal field and its uncertainty. We demonstrate the
  validity of this method by using it on a simulated galaxy catalog based on an N-body
  simulation. In doing so, we also show that the statistics of the recovered tidal field
  agree well with the Gaussian prediction (see Appendix \ref{app:th_gauss}) on large scales,
  although this agreement breaks down, as expected, on non-linear scales.

  We then apply this method to the 2MASS survey and produce a full-sky map of the transverse
  tidal field, which we make publicly
  available\footnote{\url{http://intensitymapping.physics.ox.ac.uk/2mass_tidal.html}}. The
  statistics of the
  recovered tidal field are found to agree quantitatively with our HOD-based simulated catalog
  and qualitatively with the Gaussian prediction.

  Using the recovered tidal field we identify three different environment types (knots,
  nexuses and voids) based on the eigenvalues of the tidal tensor, and compute the luminosity
  function of galaxies located in each of them. We obtain statistically significant differences
  in the bright end of the LFs, finding an excess of luminous galaxies in knots with respect
  to voids. However, we show that this effect is most likely caused by the local density of the
  environment, and not by its tidal structure. This is in agreement with previous studies
  \cite{2015MNRAS.448.3665E} and theoretical expectations \cite{2015MNRAS.447.2683A}.

  Knowledge of the tidal field also has other interesting applications in cosmology. There is
  evidence of correlations between the intrinsic shapes and alignments of galaxies
  \cite{2011JCAP...05..010B}, which can be a major source of contamination for weak lensing
  studies. If those correlations were caused by the underlying tidal forces (as suggested by
  the non-linear alignment - NLA - model \cite{2001MNRAS.320L...7C}, as well as recent
  observations \cite{2015arXiv151202236P}), prior knowledge about the
  projected tidal field could potentially reduce the effect of this systematic, effectively
  providing a prior on the contribution to the shear power spectrum from intrinsic alignments.
  Note that the projected tidal tensor (which we have shown is well approximated by the 2D tidal
  tensor) is precisely the quantity that gives rise to this contamination in the NLA model, and
  therefore the 2D tidal tensor could be used to test the validity of this model explicitly.
  
  The method presented here could therefore be beneficial in the analysis of future deeper
  photometric surveys, such as DES \cite{2005astro.ph.10346T} or LSST \cite{2009arXiv0912.0201L}.
  A possible difficulty, in this case, would be the larger projection effects of samples deeper
  than the one used here, which could significantly degrade the signal-to-noise ratio of the
  recovered tidal field. A tomographic approach with sufficiently precise photometric redshifts
  would however ameliorate these effects, and allow for a study of the statistics of the tidal
  field as a function of redshift.

\section*{Acknowledgments}
  We would like to thank Maciej Bilicki, Elisa Chisari, Thibaut Louis and Sigurd N\ae ss
  and for useful comments and discussions. DA is supported by BIPAC, ERC grant 259505 and the
  Oxford Martin School, and is grateful for the hospitality of Princeton University. BH
  acknowledges support from the Oxford Astrophysics Summer Student Programme.
  
\setlength{\bibhang}{2.0em}
\setlength\labelwidth{0.0em}
\bibliography{paper}

\appendix
\section{Spin-$s$ functions on the sphere}\label{app:th_spin}
  This section introduces a number of mathematical relations regarding
  spin-$s$ functions in $S^2$ that will be useful in Appendices \ref{app:th_gauss}
  and \ref{app:th_fullsky}.

  Let us consider the unit 2-sphere embedded in $\mathbb{R}^3$ and parametrized by the
  spherical coordinates $(\theta,\varphi)$ as ${\bf x}=(\sin\theta\,\cos\phi,
  \sin\theta\,\sin\phi,\cos\theta)$. A rotation by an angle $\psi\in[0,2\pi)$ around
  a point $\nv$ on the sphere is defined as a coordinate transformation such that
  directional vectors of the new coordinates $(\theta',\varphi')$ are rotated with
  respect to the old ones by an angle $\psi$ within the tangent plane at $\nv$.
  
  Consider now a complex function defined on the unit sphere $f(\nv)$. We say that
  $f$ is a \emph{spin-$s$ function} if its transformation law under rotations is
  $f\rightarrow f'=e^{is\psi}f$. Let us now define the so-called spin-raising and
  spin-lowering differential operators, $\eth$ and $\bar{\eth}$ respectively. They
  are defined in terms of their actions on a spin-$s$ function $_sf$:
  \begin{align}
    \eth\,_sf\equiv
       -(\sin\theta)^s(\partial_\theta+i\partial_\varphi/\sin\theta)
       (\sin\theta)^{-s}\,(_sf)\\
    \bar{\eth}\,_sf\equiv
       -(\sin\theta)^{-s}(\partial_\theta-i\partial_\varphi/\sin\theta)
       (\sin\theta)^s\,(_sf).
  \end{align}
  It is possible to prove that, if $f$ is a spin-$s$ function, then $\eth f$ and
  $\bar{\eth} f$ will be spin-$s+1$ and spin-$s-1$ quantities respectively
  \citep{spinhar}.
  
  The simplest functions we can define on the sphere are scalar (spin-$0$)
  functions. Such functions can always be expanded in terms of the ordinary spherical
  harmonics:
  \begin{equation}
    f(\nv)=\sum_{\ell=0}^\infty\sum_{m=-\ell}^{\ell}f_{\ell m}\,Y_{\ell m}(\nv).
  \end{equation}
  By applying the spin-raising and lowering operators we can define the so-called
  spin-weighted spherical harmonics $_sY_{\ell m}$, defined as
  \begin{equation}\label{eq:swsh}
    _sY_{\ell m}\equiv\left\{
    \begin{array}{cc}
      \sqrt{\frac{(\ell-s)!}{(\ell+s)!}}(\eth^s)Y_{\ell m} & 0\leq s\leq\ell \\
      (-1)^s\sqrt{\frac{(\ell+s)!}{(\ell-s)!}}(\bar{\eth}^{-s})Y_{\ell m} & -\ell\leq s\leq0 \\
      0 & \text{otherwise}
    \end{array}\right.
  \end{equation}
  Spin-$s$ functions are then amenable to a harmonic expansion in terms of the
  spin-$s$ spherical harmonics $_sY_{\ell m}$.

  The spin-weighted spherical harmonics are related to the Wigner-$d$ rotation
  matrices through:
  \begin{equation}
    _sY_{\ell m}(\theta,\varphi)=(-1)^m\sqrt{\frac{2\ell+1}{4\pi}}
    e^{im\varphi}d^\ell_{-m\,s}(\theta).
  \end{equation}
  The orthogonality of the Wigner-$d$ matrices ($\sum_{m=-\ell}^\ell
  d^\ell_{m s}\,(d^\ell_{m r})^*=\delta_{sr}$)
  then implies the following useful relation for the spin-weighted spherical harmonics:
  \begin{equation}\label{eq:ortho}
    \sum_{m=-\ell}^\ell|_sY_{\ell m}|^2=\frac{2\ell+1}{4\pi}.
  \end{equation}

\section{Gaussian statistics of the 2D tidal tensor}\label{app:th_gauss}
  This appendix discusses and derives the Gaussian prediction for the distribution of
  the 2D tidal field eigenvalues and the sky fraction occupied by the three different
  elements of the projected cosmic web.

  As outlined in Section \ref{ssec:th_2d}, the 2D tidal tensor is defined
  in terms of the projected galaxy overdensity as
  \begin{equation}
    \hat{t}\equiv\hat{H}(\nabla^{-2}_{\nv}\delta),
  \end{equation}
  where $\nabla^{-2}_{\nv}f$ denotes the particular solution of Poisson's
  equation on the 2-sphere with $f$ as a source, and $\hat{H}$ is the covariant Hessian
  defined in Eq. \ref{eq:cov_hess}. We have further defined the 2D potential
  $\phi\equiv\nabla^{-2}_{\nv}\delta$, so that $\hat{t}=\hat{H}\phi$.

  The 2D tidal tensor can also be written in closed form using the
  spin-raising and lowering operators $\eth$ and $\bar{\eth}$, introduced in the previous
  section, as
  \begin{equation}
    \hat{H}\phi\equiv\frac{1}{2}\left(
    \begin{array}{cc}
      \bar{\eth}\eth\phi+{\rm Re}(\eth\eth\phi) &
      {\rm Im}(\eth\eth\phi)\\
      {\rm Im}(\eth\eth\phi) &
      \bar{\eth}\eth\phi -{\rm Re}(\eth\eth\phi)
    \end{array}\right).
  \end{equation}
  The 2D tidal tensor can also be expressed in terms of the covariant derivatives on the
  sphere as
  \begin{equation}
    H_{ab}\phi=\frac{\nabla_a\nabla_b\phi}{|{\bf e}_a||{\bf e}_b|},
  \end{equation}
  where $\nabla_a$ denotes the covariant derivative with respect to the coordinate $q_a$
  (either $\theta$ or $\varphi$), and ${\bf e}_a\equiv\partial\nv/\partial q_a$ is the
  Vielbein of the 2-sphere.

  Let us focus now on describing the 1-point statistics of the 2D tidal tensor under the
  assumption that the underlying overdensity field is Gaussian. In this case, the probability
  distribution for $\hat{t}$ will be completely determined by the covariance of its elements:
  \begin{equation}
    C_{abcd}\equiv\langle t_{ab}t_{cd}\rangle=
    \frac{\langle \nabla_a\nabla_b\phi\,\nabla_c\nabla_d\phi\rangle}
      {|{\bf e}_a||{\bf e}_b||{\bf e}_c||{\bf e}_d|}.
  \end{equation}
  Using the isotropy of the underlying overdensity field, as well as the symmetry properties of
  the indices ($a\leftrightarrow b$, $c\leftrightarrow d$, $(a,b)\leftrightarrow (c,d)$), we can
  argue that the tensor in the numerator must be a linear combination of the only three 4-index
  isotropic tensors with equivalent symmetries:
  \begin{equation}
    \langle \nabla_a\nabla_b\phi\,\nabla_c\nabla_d\phi\rangle=
    \alpha\,g_{ab}g_{cd}+\beta\,g_{ac}g_{bd}+\gamma\,g_{ad}g_{cb},
  \end{equation}
  where $g_{ab}$ is the metric of the 2-sphere.

  Multiplying this expression above by $g^{ab}g^{cd}$, $g^{ac}g^{bd}$ and $g^{ad}g^{cb}$, and
  summing over all indices yields a linear system of three equations for the three unknown
  coefficients $\alpha$, $\beta$ and $\gamma$. Solving this system, we find that the
  covariance matrix can be written as
  \begin{align}\nonumber
    C_{abcd}=\frac{1}{8}[&(3S_A-2S_B)\delta_{ab}\delta_{cd}+\\
    &(2S_B-S_A)(\delta_{ac}\delta_{bd}+\delta_{ad}\delta_{cb})],
  \end{align}
  where $S_A$ and $S_B$ are the only two second-order rotational invariants of $\hat{t}$:
  \begin{align}
    &S_A\equiv\langle[{\rm Tr}(\hat{t})]^2\rangle=
    \langle|\bar{\eth}\eth\phi|^2\rangle,\\
    &S_B\equiv\langle{\rm Tr}(\hat{t}^2)\rangle=
    \frac{1}{2}\left(\langle|\bar{\eth}\eth\phi|^2\rangle+\langle
    |\eth\eth\phi|^2\rangle\right).
  \end{align}

  The two ensemble averages $\langle|\bar{\eth}\eth\phi|^2\rangle$ and
  $\langle|\bar{\eth}\eth\phi|^2\rangle$ can be computed using the harmonic expansion of
  $\phi$:
  \begin{equation}
    \phi(\nv)\equiv
     \sum_{\ell=0}^\infty\sum_{m=-\ell}^{\ell}\phi_{\ell m}\,Y_{\ell m}(\nv).
  \end{equation}
  Using the definition of the spin-weighted spherical harmonics introduced in Appendix
  \ref{app:th_spin} we obtain:
  \begin{align}
    &\langle|\bar{\eth}\eth\phi|^2\rangle=\sum_{\ell=0}^\infty
    C^{\phi}_\ell\left(\frac{(\ell+1)!}{(\ell-1)!}\right)^2
    \sum_{m=-\ell}^\ell|_0Y_{\ell m}|^2,\\
    &\langle|\eth\eth\phi|^2\rangle=\sum_{\ell=0}^\infty
    C^{\phi}_\ell\frac{(\ell+2)!}{(\ell-2)!}\sum_{m=-\ell}^\ell|_2Y_{\ell m}|^2
  \end{align}
  where we have defined the power spectrum of the 2D potential
  $\langle \phi_{\ell m}\phi^*_{\ell' m'}\rangle\equiv\,C^\phi_\ell
  \delta_{\ell\ell'}\delta_{mm'}$. Using the relation between the harmonic coefficients of
  the 2D potential and the projected density field ($\delta_{\ell m}=-\ell(\ell+1)
  \phi_{\ell m}$) as well as the orthogonality relation (Eq. \ref{eq:ortho}) we finally
  obtain:
  \begin{align}
    &\sigma_\delta^2\equiv\langle|\bar{\eth}\eth\phi|^2\rangle=
    \sum_{\ell=0}^\infty\frac{2\ell+1}{4\pi}C^{\delta}_\ell,\\
    &\tilde{\sigma}_\delta^2\equiv\langle|\eth\eth\phi|^2\rangle=
    \sum_{\ell=0}^\infty\frac{2\ell+1}{4\pi}
    C^{\delta}_\ell\left[\frac{(\ell+2)(\ell-1)}{\ell(\ell+1)}\right].
  \end{align}
  Note that $\sigma_\delta^2$ is the variance of the projected overdensity field. Furthermore,
  in the flat-sky approximation, where only the highest multipoles ($\ell\gg1$) contribute to
  the total power, $\tilde{\sigma}_\delta\simeq\sigma_\delta$, and thus
  $S_B\simeq S_A=\sigma_\delta^2$.

  Collecting the three independent terms of the 2D tidal tensor into a three-dimensional
  vector: ${\bf t}\equiv(t_{\theta\theta},t_{\phi\phi},t_{\theta\phi})$,
  the covariance matrix can be written as a $3\times3$ symmetric matrix in terms
  of $\sigma_\delta^2$ and $\tilde{\sigma}_\delta^2$:
  \begin{equation}
    \hat{C}\equiv\langle{\bf t}\,{\bf t}^T\rangle=\frac{1}{8}\left(
    \begin{array}{ccc}
      2\sigma_\delta^2+\tilde{\sigma}_\delta^2 & 2\sigma_\delta^2-\tilde{\sigma}_\delta^2 & 0\\
      2\sigma_\delta^2-\tilde{\sigma}_\delta^2 & 2\sigma_\delta^2+\tilde{\sigma}_\delta^2 & 0\\
      0 & 0 & \tilde{\sigma}_\delta^2
    \end{array}\right).
  \end{equation}

  In order to derive the probability distribution for the eigenvalues of $\hat{t}$, we
  follow the procedure used in \cite{2015MNRAS.447.2683A}. The probability distribution for
  the vector ${\bf t}$ is given by a multivariate Gaussian:
  \begin{equation}
    p({\bf t})\prod_{A=1}^3dt_A=\frac{\exp\left[-\frac{1}{2}{\bf t}^T\hat{C}^{-1}{\bf t}\right]}
    {\sqrt{(2\pi)^3\det(\hat{C})}}\prod_A dt_A.
  \end{equation}
  This can be simplified by defining the rescaled variables:
  \begin{equation}
    \nu\equiv\frac{t_1+t_2}{\sigma_\delta},\hspace{6pt}
    \rho\equiv\frac{t_1-t_2}{2\tilde{\sigma}_\delta},\hspace{6pt}
    \tau\equiv\frac{t_3}{\tilde{\sigma}_\delta},
  \end{equation}
  which diagonalize the covariance matrix, yielding
  \begin{equation}
    p({\bf t})\prod_{A=1}^3dt_A=\frac{8\,e^{-(\nu^2+8\rho^2+8\tau^2)/2}}
    {(2\pi)^{3/2}\sigma_\delta\tilde{\sigma}_\delta^2}\prod_A dt_A.
  \end{equation}
  On the other hand, the volume element $\prod_A dt_A$ in the space of $2\times2$ symmetric
  matrices can be written in terms of their two eigenvalues, $\lambda_1$ and $\lambda_2$
  and the angle defining the two-dimensional rotation that diagonalizes it
  \cite{1986ApJ...304...15B}:
  \begin{equation}
    \prod_A dt_A=|\lambda_1-\lambda_2|\,d\lambda_1d\lambda_2\frac{d\phi}{4},
  \end{equation}
  where the factor of $1/4$ accounts for the two different orderings of the eigenvalues and the
  overall sign defines the orientation of the eigenvectors. Chosing the ordering $\lambda_1>
  \lambda_2$ introduces a factor of 2. Expressing $\lambda_1$ and $\lambda_2$ in terms of $\nu$
  and $\rho$, and integrating the irrelevant angular part, we finally obtain the distribution:
  \begin{equation}\label{eq:pdf_nurho}
    p(\nu,\rho)d\nu d\rho=8\,\rho\,e^{-4\rho^2}\,d\rho\,\frac{e^{-\nu^2/2}}{\sqrt{2\pi}}\,d\nu
  \end{equation}
  
  The ordering $\lambda_1>\lambda_2$ has the effect of reducing the dynamical range of $\rho$
  to $\rho>0$, and the integration limits for $\nu$ are determined by the environment type,
  defined by the eigenvalue threshold $\lambda_{\rm th}$ and the number of eigenvalues
  above the threshold ($\alpha$). In general we can write $f_1(\alpha,\rho)<\nu-\nu_{\rm th}<
  f_2(\alpha,\rho)$, with $\nu_{\rm th}\equiv2\lambda_{\rm th}/\sigma_\delta$ and
  \begin{align}
    &f_1(\alpha,\rho)=\left\{\begin{array}{lc}
                             -\infty &
                              \alpha=0\,(\text{void})\\
                             -2\frac{\tilde{\sigma}_{\delta}}{\sigma_\delta}\rho & 
                              \alpha=1\,(\text{nexus}) \\
                              2\frac{\tilde{\sigma}_{\delta}}{\sigma_\delta}\rho & 
                              \alpha=2\,(\text{knot})
                            \end{array}\right.,
    \\
    &f_2(\alpha,\rho)=\left\{\begin{array}{lc}
                             -2\frac{\tilde{\sigma}_{\delta}}{\sigma_\delta}\rho &
                              \alpha=0\,(\text{void}) \\
                              2\frac{\tilde{\sigma}_{\delta}}{\sigma_\delta}\rho &
                              \alpha=1\,(\text{nexus})\\
                              \infty &
                              \alpha=2\,(\text{knot})
                            \end{array}\right..
  \end{align}
  The sky fraction for the three types of environment can then be computed by integrating
  the probability distribution in Eq. \ref{eq:pdf_nurho} with the corresponding integration
  limits for $\nu$:
  \begin{equation}
    F_V(\alpha,\lambda_{\rm th})=\frac{8}{\sqrt{2\pi}}\int_0^\infty d\rho \int_{\nu_{\rm th}+
    f_1(\alpha,\rho)}^{\nu_{\rm th}+f_2(\alpha,\rho)}d\nu
    \,\rho\,e^{-(\nu^2+8\rho^2)/2}.
  \end{equation}
  Note that this integral can be solved analytically in terms of error functions, but we omit
  the resulting cumbersome expression.

\section{Relation between the 2D tidal tensor and the projected tidal forces}
\label{app:th_fullsky}
  \begin{figure}
    \centering
    \includegraphics[width=0.5\textwidth]{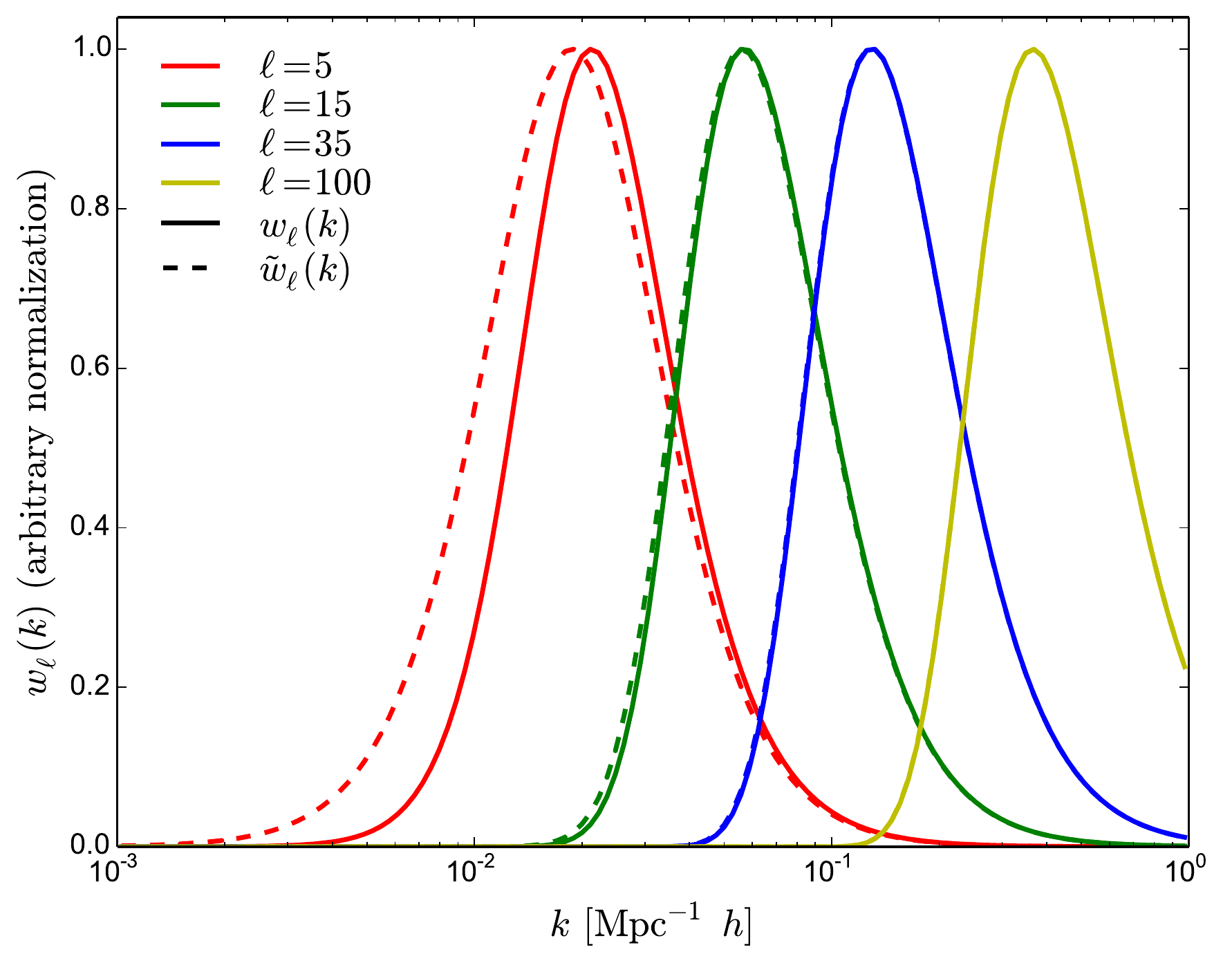}
    \caption{Window functions $w_\ell(k)$ and $\tilde{w}_\ell(k)$ defined in Equations 
             \ref{eq:wl} and \ref{eq:wlt} for the 2MASS selection function and different values
             of the multipole order $\ell$. The difference between the 2D tidal tensor and the
             projected tidal forces is effectively encapsulated in the differences between
             the two functions. These differences become negligible by $\ell\simeq15$, and
             hence, for most scales of interest, it is safe to interpret the 2D tidal tensor
             as describing the transverse tidal forces averaged along the line of sight.}
    \label{fig:windows}
  \end{figure}
  Using the flat-sky approximation, we have shown in Section \ref{sssec:th_2d_phys} that the 2D
  tidal tensor, as defined in Section \ref{sssec:th_2d_def} can be interpreted, on
  sufficiently small scales, as being proportional to the gravitational tidal forces in the
  transverse (angular) directions averaged along the line of sight with the survey selection
  function. The aim of this Appendix is to present a more rigorous proof of this relation in
  the full-sky limit, as well as to provide the formulas relating the 2D tidal tensor and the
  underlying matter perturbations.

  As shown in Appendix \ref{app:th_gauss}, the components of the 2D tidal tensor can be related
  to the two complex quantities
  \begin{align}\label{eq:def_t0t2}
    t^{(0)}(\nv)\equiv\bar{\eth}\eth\left[\nabla^{-2}_{\nv}
    \left(\int_0^\infty d\chi\,w(\chi)\,\delta^s(\chi\nv)\right)\right],\\
    t^{(2)}(\nv)\equiv\eth\eth\left[\nabla^{-2}_{\nv}
    \left(\int_0^\infty d\chi\,w(\chi)\,\delta^s(\chi\nv)\right)\right],
  \end{align}
  where $\delta^s({\bf x})$ is the observed density field in redshift space. On the other hand,
  the projected tidal tensor (i.e. the tidal forces along the transverse directions averaged
  along the line of sight) is given by the analogous quantities:
  \begin{align}
    &\tilde{t}^{(0)}(\nv)\equiv\int_0^\infty d\chi\,w(\chi)
    \frac{\bar{\eth}\eth\Phi(\chi\nv)}{\chi^2},
    \\
    &\tilde{t}^{(2)}(\nv)\equiv\int_0^\infty d\chi\,w(\chi)
    \frac{\eth\eth\Phi(\chi\nv)}{\chi^2},
  \end{align}
  where $\Phi$ is the Newtonian potential normalized so that $\nabla^2\Phi=\delta$, with
  $\delta$ the real-space density field. Let us focus, for the moment, on the trace of the
  two tidal tensors, given by $t^{(0)}$ and $\tilde{t}^{(0)}$.

  We start by computing the harmonic coefficients of both quantities, given by
  \begin{equation}
    t^{(0)}_{\ell m}\equiv\int d\nv\,Y^*_{\ell m}(\nv)\,t^{(0)}(\nv),
  \end{equation}
  (and likewise for $\tilde{t}^{(0)}$). We can relate $t^{(0)}_{\ell m}$ to the matter
  density perturbations by doing the following:
  \begin{itemize}
    \item First, expand $\delta_s$ in Eq. \ref{eq:def_t0t2} in terms of its Fourier
          coefficients, and use the plane-wave expansion
          \begin{equation}
            e^{i{\bf k}{\bf x}}=\sum_{\ell=0}^\infty 4\pi i^\ell j_\ell(k\chi)\,
            \sum_{m=-\ell}^{\ell}Y_{\ell m}(\nv)Y^*_{\ell m}(\nv_k),
          \end{equation}
          where ${\bf x}\equiv\chi\nv$, $\nv_k$ is the unit vector in Fourier space
          and $j_\ell(x)$ is the order-$\ell$ spherical Bessel function of the first kind.
    \item Apply the operator $\bar{\eth}\eth\nabla^{-2}_{\nv}$ on the spherical harmonic
          $Y_{\ell m}(\nv)$
          \begin{equation}
            \bar{\eth}\eth\nabla^{-2}_{\nv}Y_{\ell m}(\nv)=Y_{\ell m}(\nv)
          \end{equation}
    \item Relate the Fourier coefficients of the observed overdensity field $\delta^s$
          to the real-space density perturbations:
          \begin{equation}
            \delta^s_{\bf k}\,j_\ell(k\chi)\longrightarrow b\,\delta_{\bf k}
            \left[j_\ell(k\chi)-\beta j_\ell''(k\chi)\right],
          \end{equation}
          where $b$ is the linear galaxy bias and $\beta$ is the redshift distortion
          parameter.
    \item Define the harmonic coefficients of $\delta_{\bf k}$ as
          \begin{equation}
            \delta_{\ell m}(k)\equiv\int d\nv_k Y^*_{\ell m}(\nv_k)\,\delta_{\bf k}.
          \end{equation}
  \end{itemize}
  Finally we obtain the following relation between $t^{(0)}_{\ell m}$ and $\delta_{\ell m}(k)$:
  \begin{equation}\label{eq:wl}
    t^{(0)}_{\ell m}= b\,\frac{4\pi i^\ell}{(2\pi)^{3/2}}
    \int_0^\infty dk\,k^2\,\delta_{\ell m}\,w_\ell(k),
  \end{equation}
  with
  \begin{equation}
    w_\ell(k)\equiv\int_0^\infty d\chi\,
    \left[j_\ell(k\chi)-\beta j_\ell''(k\chi)\right]w(\chi).
  \end{equation}

  Following analogous steps for $\tilde{t}^{(0)}_{\ell m}$ yields a similar relation:
  \begin{equation}\label{eq:wlt}
    \tilde{t}^{(0)}_{\ell m}=\frac{4\pi i^\ell}{(2\pi)^{3/2}}
    \int_0^\infty dk\,k^2\,\delta_{\ell m}\,\tilde{w}_\ell(k),
  \end{equation}
  with
  \begin{equation}
    \tilde{w}_\ell(k)\equiv\int_0^\infty d\chi\,
    \frac{\ell(\ell+1)}{(k\chi)^2}\,j_\ell(k\chi)\,w(\chi).
  \end{equation}
  Thus, as was shown in the flat-sky case, the differences between the 2D tidal tensor and
  the projected tidal forces, besides the linear galaxy bias acting as a proportionality
  constant, are encapsulated in the different window functions $w_\ell$ and
  $\tilde{w}_\ell$ above.

  Figure \ref{fig:windows} shows both window functions for different values of $\ell$ for the
  2MASS selection function using an RSD parameter $\beta=0.46$ \citep{2015MNRAS.449..670A}. As
  is evident, on small scales (large-$\ell$) both functions are almost equivalent, and in this
  case the 2D tidal tensor can be safely interpreted as describing the average tidal forces in
  the transverse directions.

  Similar relations can be derived for $t^{(2)}$ and $\tilde{t}^{(2)}$ by following the same
  steps outlined above, with the exception that, since they are spin-$2$ quantities, their
  harmonic coefficients must be computed using the spin-weighed spherical harmonics
  (Eq. \ref{eq:swsh}). The resulting espressions are
  \begin{align}
    &t^{(2)}_{\ell m}= b\,\frac{4\pi i^\ell}{(2\pi)^{3/2}}
    \int_0^\infty dk\,k^2\,\delta_{\ell m}\,w^{(2)}_\ell(k),\\
    &\tilde{t}^{(2)}_{\ell m}=\frac{4\pi i^\ell}{(2\pi)^{3/2}}
    \int_0^\infty dk\,k^2\,\delta_{\ell m}\,\tilde{w}^{(2)}_\ell(k),
  \end{align}
  with
  \begin{align}
    &\binom{w^{(2)}_\ell(k)}{\tilde{w}^{(2)}_\ell(k)}\equiv
    \left[\frac{(\ell+2)(\ell-1)}{\ell(\ell+1)}\right]^{1/2}
    \binom{w_\ell(k)}{\tilde{w}_\ell(k)}.
  \end{align}

\end{document}